\documentclass[twocolumn,numberedappendix]{emulateapj}
\bibpunct[; ]{(}{)}{;}{a}{}{;}
\usepackage{apjfonts}

\usepackage{graphicx} 
\usepackage{natbib}

\newcommand{\Msolar}{M${_\odot}$\,}

\shorttitle{Photometric determination of the mass accretion rates of
pre-main sequence stars in the SN\,1987A field}
\shortauthors{De Marchi, Panagia \& Romaniello}

\slugcomment{Accepted for publication in ``The Astrophysical Journal''}

\begin{document}

\title{Photometric determination of the mass accretion rates of
pre-main sequence stars. I. \\ Method and application to the SN\,1987A
field\footnote{\small Based on observations with the NASA/ESA {\it Hubble Space
Telescope}, obtained at the Space Telescope Science Institute, which is
operated by AURA, Inc., under NASA contract NAS5-26555}}


\author{
Guido De Marchi,\altaffilmark{1}
Nino Panagia,\altaffilmark{2,3,4} and
Martino Romaniello\altaffilmark{5}
}

\altaffiltext{1}{European Space Agency, Space Science Department, Keplerlaan
1, 2200 AG Noordwijk, Netherlands; gdemarchi@rssd.esa.int}

\altaffiltext{2}{Space Telescope Science Institute, 3700 San Martin Drive, 
Baltimore, MD 21218, USA, panagia@stsci.edu}
 
\altaffiltext{3}{INAF--CT, Osservatorio Astrofisico di Catania, Via S. Sofia 
78, 95123 Catania, Italy}

\altaffiltext{4}{Supernova Limited, VGV \#131, Northsound Road, Virgin Gorda, 
British Virgin Islands}

\altaffiltext{5}{European Southern Observatory, Karl--Schwarzschild-Str. 2, 
85748 Garching, Germany}

\begin{abstract}  

We have developed and successfully tested a new self-consistent method
to reliably identify pre-main sequence (PMS) objects actively
undergoing mass accretion in a resolved stellar population, regardless
of their age. The method does not require spectroscopy and combines
broad-band $V$ and $I$ photometry with narrow-band $H\alpha$ imaging
to: (1) identify all stars with excess H$\alpha$ emission; (2) convert
the excess H$\alpha$ magnitude into H$\alpha$ luminosity $L(H\alpha)$;
(3) estimate the H$\alpha$ emission equivalent width; (4) derive the
accretion luminosity $L_{\rm acc}$ from $L(H\alpha)$; and finally (5)
obtain the mass accretion rate $\dot M_{\rm acc}$ from $L_{\rm acc}$
and the stellar parameters (mass and radius). By selecting stars with
an accuracy of 15\,\% or better in the H$\alpha$ photometry, the
statistical uncertainty on the derived $\dot M_{\rm acc}$ is typically
$\lesssim 17\,\%$ and is dictated by the precision of the H$\alpha$
photometry. Systematic uncertainties, of up to a factor of 3 on the
value of $\dot M_{\rm acc}$, are caused by our incomplete understanding
of the physics of the accretion process and affect all determinations
of the mass accretion rate, including those based on a spectroscopic
H$\alpha$ line analysis.

As an application of our method, we study the accretion process in a
field of $9.16$ arcmin$^2$ around SN\,1987A, using existing Hubble
Space Telescope photometry. We identify as bona-fide PMS stars a total
of 133 objects with a H$\alpha$ excess above the $4\,\sigma$ level and a
median age of $13.5$\,Myr. Their median mass accretion rate of $2.6
\times 10^{-8}$\,\Msolar\,yr$^{-1}$ is in excellent agreement with
previous determinations based on the $U$-band excess of the stars in
the same field, as well as with the value measured for G-type PMS stars
in the Milky Way. The accretion luminosity of these PMS objects shows a
strong dependence on their distance from a group of hot massive stars
in the field and suggests that the ultraviolet radiation of the latter
is rapidly eroding the circumstellar discs around PMS stars.

\end{abstract}


\section{Introduction}

In the current star formation paradigm, low-mass stars grow in mass
over time through accretion of matter from a circumstellar disc (e.g.
Lynden-Bell \& Pringle 1974; Appenzeller \& Mundt 1989; Bertout 1989).
This accretion process is believed to occur via the magnetic field
lines that connect the star to its disc and that act as funnels for the
gas (e.g. K\"onigl 1991; Shu et al. 1994). The strong  excess emission
observed in most of these pre-main sequence (PMS) stars is believed to
originate through gravitational energy, released by infalling matter,
that ionises and excites the gas. Thus, the excess luminosity of
these objects  (the classical T Tauri stars) can be used to measure
the mass accretion rate. 

A reliable measurement of the rate of mass accretion onto PMS stars is
of paramount importance for understanding the evolution of both the
stars and their discs (see e.g. review by Calvet et al. 2000). Of
particular interest is determining how the mass accretion rate changes
with time as a star approaches its main sequence, whether it
depends on the final mass of the forming star and whether it is
affected by the chemical composition and density of its parent molecular
cloud or by the proximity of hot, massive stars. 

Ground-based spectroscopic studies of nearby young star-forming regions
(e.g. in Taurus, Auriga, Ophiuchus, Orion) show that the mass
accretion  rate appears to decrease steadily with time, from $\sim
10^{-8}$\,\Msolar\,yr$^{-1}$ at ages of $\sim 1$\,Myr to $<
10^{-9}$\,\Msolar\,yr$^{-1}$ at  $\sim 10$\,Myr (Muzerolle et al. 2000;
Sicilia-Aguilar et al. 2005; 2006). While at face value this is in line
with the expected evolution of viscous discs (Hartmann et al. 1998),
the scatter on the data is very large, exceeding 2~dex at any given
age. This is at least in part explained by the recent finding that the
mass accretion rate seems also to depend quite steeply on the mass of
the forming star. Muzerolle et al. (2003; 2005), Natta et al. (2004;
2006), White \& Hillenbrand (2004) and Sicilia-Aguilar et al. (2006)
consistently report that the accretion rate $\dot M_{\rm acc}$
decreases with the stellar mass $M$ as $\dot M_{\rm acc} \propto M^2$,
albeit with large uncertainties on the actual value of the power-law
index. On the other hand, Clarke \& Pringle (2006) warn that such a
steep decline might be spurious and strongly driven by
detection/selection thresholds.

While observational uncertainties and potential systematic errors may
well contribute to the present uncertainty, the true limitation in this
field of research currently comes from the paucity of available
measurements. Indeed, all the results so far obtained are based on the
mass accretion rates of a small number of stars (just over one
hundred), all located in nearby Galactic star forming regions, covering
a very limited range of ages and no appreciable range of metallicity
(all clouds  mentioned above having essentially solar metallicity; e.g.
Padgett 1996).

The origin of this limitation is to be found in the technique used to
measure PMS mass accretion rates. Regardless as to whether the latter
are derived from the analysis of veiling in the photosperic absorption
lines (usually at ultraviolet and optical wavelengths) or through a
detailed study of the profile and intensity of emission lines (usually
H$\alpha$, Pa$\beta$ or Br$\gamma$), these thechniques typically
require medium to high resolution spectroscopy for each individual
object. Even with modern multi-object spectrographs at the largest
ground-based telescopes, this approach is not very efficient and cannot
reach beyond nearby star forming regions or the Orion cluster, as
crowding and sensitivity issues rule out but the brightest objects in
more distant regions like the centre of the Milky Way or the Magellanic
Clouds. While in principle the  PMS mass accretion rate can be
estimated from the $U$-band excess if the  spectral type is known (see
e.g. Gullbring et al. 1998; Romaniello et al. 2004), this method is
potentially subject to large uncertainties in the extinction and in the
determination of the correct spectral type and effective temperature
when spectroscopy is not available.

In order  to overcome these limitations and significantly extend the
sample of stars with a measured mass accretion rate, we have developed
and successfully tested a new method to reliably measure $\dot M_{\rm
acc}$ that does not require spectroscopy. The method, described in the
following sections, combines broad-band $V$ and $I$ photometry with
narrow-band $H\alpha$ imaging and allows us to identify all stars with
excess H$\alpha$ emission and to derive from it the accretion
luminosity $L_{acc}$ and hence $\dot M_{\rm acc}$ for hundreds of
objects simultaneously.

The paper is organised as follows: in Section\,2 we address the
identification of PMS stars via their colour excess and in Section\,3
we show how to reliably convert the measured excess into H$\alpha$
luminosity and equivalent width. Section\,4 shows how the accretion
luminosity and mass accretion rate can be obtained from the H$\alpha$
luminosity and Section\,5 presents an application of the method to the
field of SN\,1987A, in the Large Magellanic Cloud (LMC). A thorough
discussion of all systematic uncertainties involved in the method is
provided in Section\,6, while a general discussion and conclusions
follow in Section\,7. The Appendix provides additional figures and
tabular material.

\section{Identification of PMS stars through their colour excess}

One of the characteristic signatures of the accretion process on to PMS
stars is the presence of excess emission in H$\alpha$ (see Calvet et
al. 2000). Although surveys to search for stars with H$\alpha$ excess
emission over large sky areas have been traditionally carried out via
slitless spectroscopy (e.g. Herbig 1957; Kohoutek \& Wehmeyer 1999),
extensive studies in which narrow-band H$\alpha$ imaging is combined
with broad-band photometry also exist (e.g. Parker et al. 2005; Drew et
al. 2005). 

Recently, Romaniello et al. (1998), Panagia et al. (2000) and
Romaniello et al. (2006) used HST/WFPC2 photometry in H$\alpha$ (F656N)
and the R band (F675W) to provide the first direct detection of
extra-galactic PMS stars with strong H$\alpha$ emission in the LMC.
Their approach is based on the widely adopted use of the R band
magnitude as an indicator of the level of the photospheric continuum
near the H$\alpha$ line. Therefore, stars with strong H$\alpha$
emission will have a large $R-H\alpha$ colour. From the analysis of the
specific F656N and F675W filter response curves of the WFPC2
instrument, Panagia et al. (2000) conclude that an equivalent width
($W_{\rm eq}$) of the H$\alpha$ emission line in excess of 8\,\AA\,\,
will result in a colour excess $R-H\alpha \gtrsim 0.3$.  

This way of identifying PMS stars is more accurate and reliable than
the simple classification based on the position of the objects in the
Hertzsprung--Russell diagram, i.e. stars placed well above the main
sequence (e.g. Gilmozzi et al. 1994; Hunter et al. 1995; Nota et al.
2006; Gouliermis et al. 2007). The reason is twofold: {\em (i)}
contamination by older field stars and the effects of differential
extinction may lead to an  overestimate of the actual number of
candidate young, low-mass PMS stars, i.e. those ``above'' the MS; {\em
(ii)} older PMS stars, already close to their MS, cannot be properly
identified with this method, thereby resulting in an underestimate of
the number of stars that are still forming. When PMS stars are
identified by means of their $R-H\alpha$ colour excess, both problems
disappear.

Similarly, this approach is more practical, efficient and economical
than slitless spectroscopy for studying PMS stars in dense star forming
regions in the Milky Way and nearby galaxies, which are today
accessible thanks to high-resolution observations, in particular those
made with the HST.

We will show in Section\,3 that the level of the photospheric continuum
near H$\alpha$ can be directly derived from the observed $V$ and $I$
magnitudes. This means that PMS stars with an H$\alpha$ excess can be
identified without actually measuring the canonical $R-H\alpha$ index.
On the other hand, since $R-H\alpha$ has traditionally been used in PMS
stars studies, we first show here how one can derive analytically the
$R$ magnitude of an object from the measured $V$ and I magnitudes. In
this way, we intend to make it possible for those who do not dispose of
R-band photometry for their fields to derive it in an accurate way by
using measurements in the neighbouring V and I bands. The latter
are more commonly used than the R band in photometric studies of
resolved stellar populations and are particularly popular among HST
observers.

\subsection{Deriving $R$ from $V$ and $I$}

The absence of remarkable spectral features between 5\,000\,\AA\, and
9\,000\,\AA\, in the stellar continuum of dwarfs and giants with
effective temperatures in the range 4\,000\,K $\lesssim T_{\rm
eff} \lesssim$ 10\,000\,K implies that their $V-R$ and $V-I$
colours depend primarily on the effective temperature of the stars and
less on their metallicity. In effect, $V-I$ remains a useful colour
index for temperature determinations (von Braun et al. 1998; Bessell,
Castelli \& Plez 1998).  

\begin{figure}
\centering
\resizebox{\hsize}{!}{\includegraphics[width=16cm]{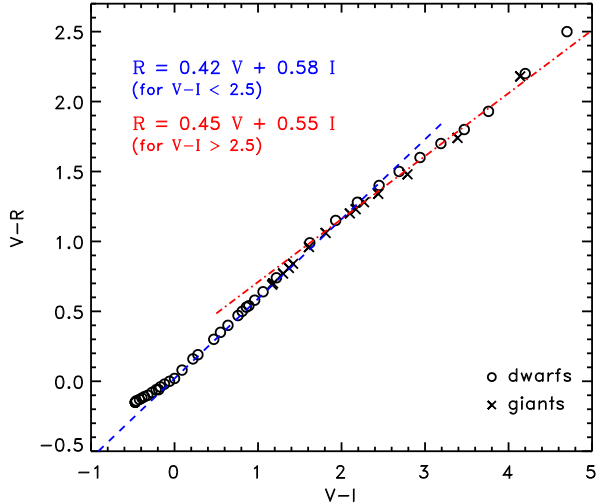}}
\caption{The tight correlation between $V-R$ and $V-I$ from the
colour--colour sequences of Johnson (1966). Circles indicate dwarfs and
crosses giants.} 
\label{fig1}
\end{figure}

The colour--colour sequences in the $UBVRIJKLMN$ passbands published by
Johnson (1966) revealed a very tight correlation between the mean $V-R$
and $V-I$ colours of local dwarfs and giants (see Figure\,\ref{fig1}).
The correlation is remarkably linear in the range $0 < V-I < 2$ (i.e.
4\,000\,K $\lesssim T_{\rm eff} \lesssim$ 10\,000\,K), where $(V-R) =
0.58 \times (V-I)$ (dashed line in Figure\,\ref{fig1}), and it remains
linear for $2 < V-I < 4$ (i.e. 3\,000\,K $\lesssim T_{\rm eff}
\lesssim$ 4\,000\,K), albeit with a shallower slope ($0.45$; dot-dashed
line). It is, therefore, possible and meaningful to attempt to derive
the value of the $R$ magnitude from the knowledge of $V$ and $I$,
particularly because no extrapolation beyond the available wavelength
range is involved. The  simple inversion of the relationship above
gives $R = 0.42 \, V + 0.58 \, I$ for the system of
Johnson.\footnote{We note here that Johnson's $I$ is not the same as
the much more commonly used Cousin's $I$. Therefore, the colour
relationships given here cannot be used for the Cousin system. Proper
relationships for the most common magnitude systems are provided in the
Appendix.}

Interestingly, this relationship is very similar to the one that we
would obtain if we assumed that the proportion with which the $V$ and
$I$ magnitudes contribute to the $R$ magnitude is simply dictated by
the ratio of the effective wavelengths of the filters. The effective
wavelengths of the  V, R and I bands in Johnson's system are,
respectively, $\lambda_e^V = 5500$\,\AA\,, $\lambda_{\rm e}^{\rm }R 
=7000$\,\AA\, and $\lambda_{\rm e}^{\rm V} = 9000$\,\AA\,, which imply
that the V band should contribute for $(7000-5500)/(9000-5500) \simeq
0.43$ and the I band for $(9000-7000)/(9000-5500) \simeq 0.57$ to the
R-band magnitude. These values are remarkably close to those observed. 

The apparent disarming simplicity of this conclusion, however, should
not lead one to think that this is the case for any photometric system.
In particular, for spectral bands much wider than Johnson's, such as
some of the HST/WFPC2 and HST/ACS filters, the matter becomes more
complicated because the true effective wavelength can depend
significantly on the spectral properties of the source (i.e. its
temperature). In these cases, the relationship should be derived via
direct observations in the specific bands or through synthetic
photometry, when the properties of the photometric system are well
caracterised. 

Due to the large number of inter-related observing modes offered by the
scientific instruments on board the HST, synthetic photometry has long
been established as the most practical and efficient approach to their 
calibration (see Koornneef et al. 1986; Horne 1988; Sirianni et al.
2005). For this reason, reference tables exist that accurately describe
the main optical components of the telescope and instruments as well as
the response and quantum efficiency of the detectors as a function of
wavelength. Therefore, thanks also to the availability of standardised
software packages (e.g. Synphot; see Laidler et al. 2008), it is
relatively easy to derive synthetic magnitudes from observed or
theoretical stellar spectra for various HST instrumental
configurations. 

We have used both observed and theoretical stellar spectra, taken
respectively from the Bruzual-Persson-Gunn-Stryker (BPGS)
Spectrophotometry Atlas (see Gunn \& Stryker 1983) and from the stellar
atmosphere models of Bessell et al. (1998), to compute colour
relationships between  $R$-like magnitudes and $V-I$-like colours for
various HST instruments and ground photometric systems. Details on how
the colour transformations were derived and a table of coefficients
(Table\,\ref{tab3}) can be found in the Appendix, while a brief
comparison with HST data is given in the following section.

\begin{figure}
\centering
\resizebox{\hsize}{!}{\includegraphics[width=16cm]{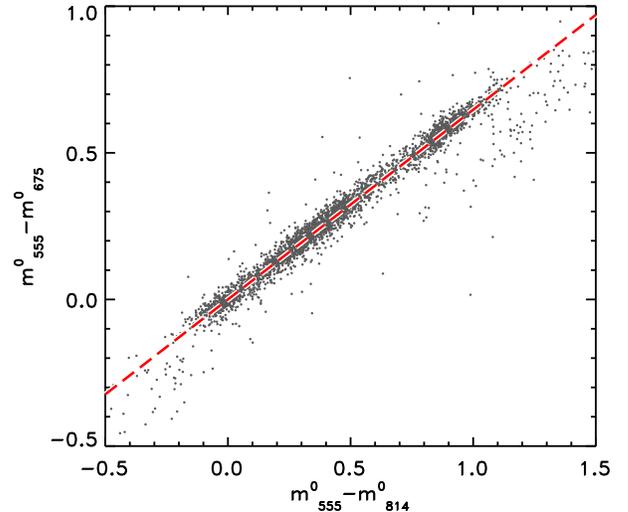}}
\caption{Colour--colour diagram of all stars in the SN\,1987A field
with combined mean photometric error $\delta_3 \le 0.05$\,mag. The
dashed line shows the colour relationship derived for the bands in
question (F555W, F675W and F814W) from the models of Bessell et al.
(1998) and is in excellent agreement with the observations.} 
\label{fig2}
\end{figure}

\begin{figure*}
\centering
\includegraphics[width=16cm]{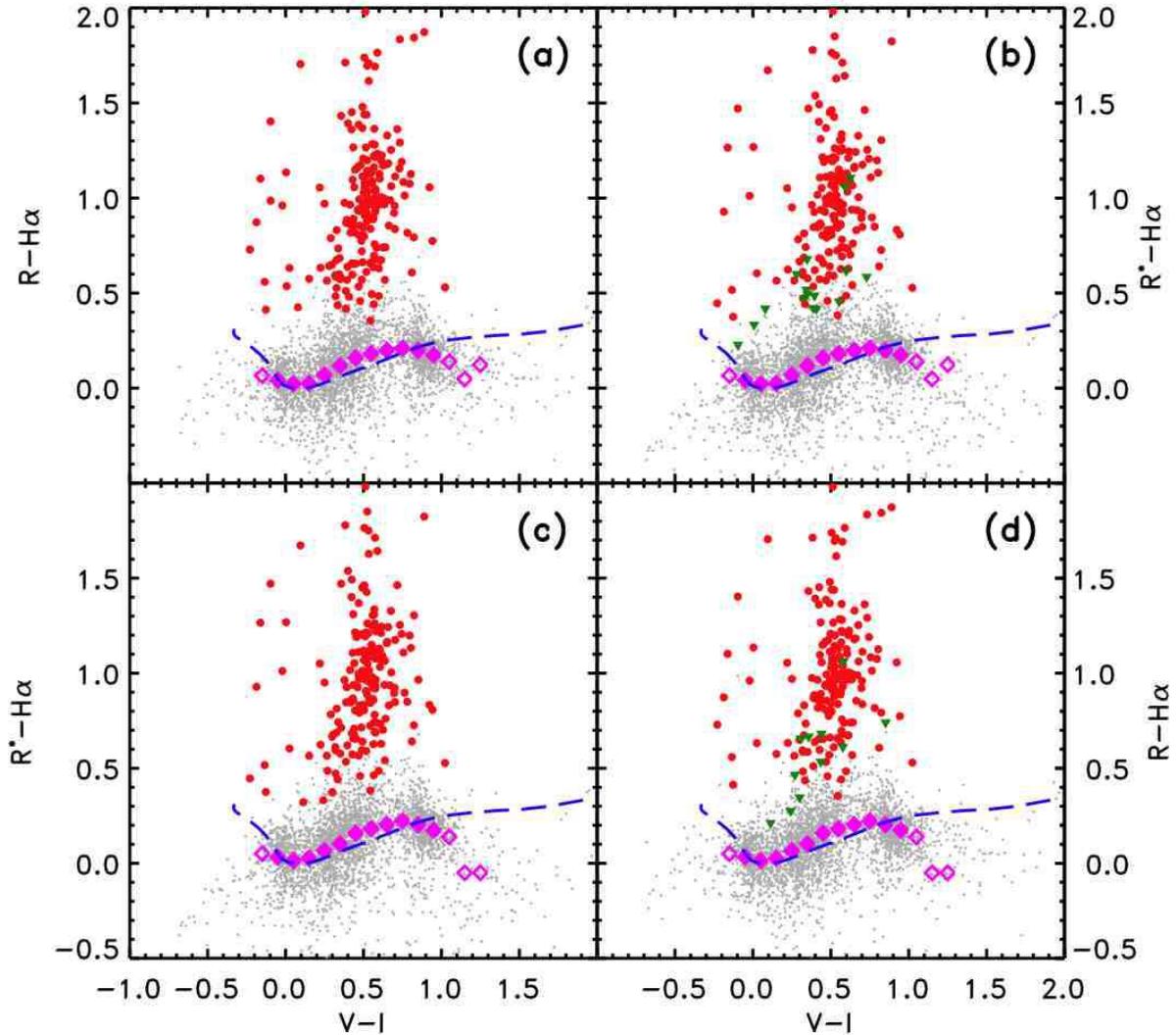}
\caption{Comparison between the measured and derived $R - H\alpha$
colour excess. While $R$ denotes the magnitude actually measured in the
F675W band, $R^*$ indicates the magnitude derived from $V$ (F555W) and
$I$ (F814W) by using the relationship of Table\,\ref{tab3}. 
Diamonds represent the measured average $R-H\alpha$ colour of the
population while the dashed lines is the colour predicted my theoretical
models. Thick dots corresponds to stars with $R-H\alpha$ excess larger
than $4\,\sigma$. See text for an explanation of the remaining symbols.}
\label{fig3}
\end{figure*}

\subsection{Comparison with the data}

To verify the validity of the predicted colour relationships, we have
used WFPC2 photometry of a field around SN\,1987A studied by Romaniello
(1998), Panagia et al. (2000) and Romaniello et al. (2002). The
observations cover a field of $9.16$\,arcmin$^2$ and were collected
over three epochs, namely 1994 September, 1996 February and 1997 July.
The advantage of this catalogue is that the magnitude of each star in
it is individually corrected for the effects of interstellar extinction
by using the information provided simultaneously by all broad-band
filters (see Romaniello et al. 2002 for more details). While it would
be possible to include the effects of reddening in the calculations of
the colour relationships of Table\,\ref{tab3}, it is more appropriate
in this case to correct the observed magnitudes, since considerable
differential extinction is likely to be present in this field (see
Panagia et al. 2000; De Marchi \& Panagia, in preparation).

From the catalogue of Romaniello et al. (2002), we have selected all
those stars whose mean error $\delta_3$ in the three bands F555W, F675W
and F814W does not exceed $0.05$\,mag, where

\begin{equation} \delta_3 =
\sqrt{\frac{\delta^2_{555}+\delta^2_{675}+\delta^2_{814}}{3}}
\label{eq1} 
\end{equation} 

\noindent 
and $\delta_{555}$, $\delta_{675}$ and $\delta_{814}$ are the
photometric uncertainties in each individual band after correction for
interstellar extinction. A total of $9\,635$ stars satisfy the
condition set by Equation\,\ref{eq1}, out of $21\,955$ objects in the
complete catalogue. 

The colour--colour diagram for the stars selected in this way is shown
in Figure\,\ref{fig2}. The dashed line in that figure corresponds to
the colour relationship derived for the bands in question from the
models of Bessell et al. (1998; see the Appendix) and is practically
indistinguishable from the best linear regression fit to the data,
which has a slope of $0.645 \pm 0.003$. Although these models were
selected to match the metallicity of the LMC, having used the
relationship derived with the BPGS atlas or that for $[M/H]=-1$ would
have still resulted in a very good fit, within the observational
uncertainties. This result confirms that, at least over the range
explored here, metallicity does not play a dominant role in the colour
relationships. 

We provide a further proof of the reliability and accuracy of the
interpolated R-band magnitudes in Figure\,\ref{fig3}, where we compare
the identification of PMS stars based on the measured and interpolated
$R-H\alpha$ colour. To distinguish the observed from  interpolated
R-band magnitude, we indicate the latter with $R^*$, as explained
below.\footnote{Hereafter, we  will use for simplicity the symbols $V$,
$R$, $I$ and $H\alpha$ to indicate the respective WFPC2 bands. When
differences between photometric systems are important (e.g. in
Table\,\ref{tab3}), all bands will be indicated with their specific
name.} In this case, from the photometry of Romaniello et al. (2002) 
already corrected for reddening, we have further restricted the
selection to objects with $\delta_3 \le 0.02$, comprising $3\,241$
stars. 

Diamonds in Figure\,\ref{fig3} represent the average $R-H\alpha$ colour
of the population and, as such, define the reference with respect to
which one should look for excess emission. The dashed lines represent
the theoretical colour relationship obtained using the Bessell et al.
(1998) model atmospheres mentioned above for normal main-sequence stars
and is in rather good agreement with the average observed colours in
the range $-0.1 \lesssim V-I \lesssim 1$ (filled diamonds).
The rms deviation between the model and the data over this range
amounts to $0.03$\,mag and is dominated by the systematic departure in
the range $0.2 \lesssim V-I \lesssim 0.7$. The latter could
be due to a number of reasons, including the inability of the models to
accurately describe the properties of the H$\alpha$ line since they are
meant to be used for broad-band photometry. The departure of the models
from the data at $V-I > 1$ stems from both small number
statistics and the fact that the bulk of the population there comes
from red giants, which have lower surface gravity and possibly lower
metallicity than those assumed in the models. In the following, we will
concentrate our analysis to the range $-0.1 \lesssim  V-I
\lesssim 1$ marked by the filled diamonds. 

Figure\,\ref{fig3}a is based on the observed data and we indicate as
large dots all stars whose observed $R-H\alpha$ colour exceeds the
local average by at least four times the individual photometric
uncertainty in the colour, i.e. the one specific to each star. 
Consequently, these objects must be regarded as having a bona-fide
H$\alpha$ excess above the $4\,\sigma$ level. There are 199 such
objects in panel (a). In panel (b) we have replaced the observed $R$
magnitude with its interpolated value $R^*$. All but 16 of the 199
stars with $4\,\sigma$ excess in panel (a) have a $4\,\sigma$ excess
also in panel (b) and are indicated as large dots. The remaining 16
objects, marked as triangles, fail this requirement as their colour is
closer to the local average colour than four times their specific
combined photometric uncertainty.

In panels (c) and (d) we study the uncertainty in the other direction,
i.e. from the interpolated to the observed plane. Panel (c) simulates
the case in which, no $R$-band data being available, one uses the 
interpolated $R^*$ value from Table\,\ref{tab3} to identify stars with
H$\alpha$ excess emission. Stars indicated as large dots in panel (c)
are those with interpolated $R^*-H\alpha$ colour exceeding the local
average by at least four times the individual photometric uncertainty
in the colour, i.e. those with H$\alpha$ excess above the $4\,\sigma$
level. A total of 194 stars meet this condition. All but 11 of them
would have met the $4\,\sigma$ excess condition if we had used the
actually observed $R$-band magnitude. They are indicated as tlarge dots
in panel (d), while the 11 objects failing the condition are marked as
triangles.

From Figure\,\ref{fig3} we can draw some quantitative conclusions on 
the statistical significance of the detection of the colour excess in
H$\alpha$. First of all, it does not surprise that some objects
detected as bona-fide H$\alpha$ excess stars in panel (a) are rejected
in panel (b), since the typical photometric uncertainty on $R^*$
is a factor of $\sqrt{2}$ higher than that on $R$, as one must
combine the uncertainties on $V$ and $I$ that concur to the
estimate of $R^*$ (although the major contribution to the
uncertainty on the $R-H\alpha$ colour of a star comes from the
H$\alpha$ band). For the same reason, it is to be expected that panel
(c) will yield less objects with bona-fide H$\alpha$ excess than panel
(a). 

Considering the number of detections and mismatches of
Figure\,\ref{fig3}, we estimate that using  the interpolated $R^*$ in
place of $R$ a fraction of $(199-194)/199 = 2.5\,\%$ of the objects with
H$\alpha$ excess at the $4\,\sigma$ level would be missing and that for
about $11/194=5.7\,\%$ of the detections the assigned statistical
significance would be higher than indicated by direct $R$ measurements.
We conclude that our interpolation scheme is quite reliable in that it
can provide the correct identification of PMS stars in no less 94\,\%
of the cases.

\section{Quantifying the excess H$\alpha$ emission: From colour excess
to line luminosity and equivalent width}

While helpful to accurately identify PMS stars, the $R-H\alpha$ colour
excess alone does not immediately provide an absolute measure of the
H$\alpha$ luminosity $L(H\alpha)$ nor of the equivalent width $W_{\rm
eq}(H\alpha)$, which are necessary to arrive at the mass accretion
rate. In order to measure $L(H\alpha)$, one needs a solid estimate of
the stellar spectrum in the H$\alpha$ band (i.e. without the
contribution of the emission). Similarly, to measure $W_{\rm
eq}(H\alpha)$, a solid estimate of the level of the stellar continuum
inside the H$\alpha$ band is needed. Since the R band is over an order
of magnitude wider than the H$\alpha$ filter, it cannot provide this
information accurately, but only gives a rough estimate of the
continuum level. In this section we present a simple method that allows
us to measure $L(H\alpha)$ for the objects with an H$\alpha$ excess and
to derive the level of their spectral continuum inside the H$\alpha$
band, determining in this way also $W_{\rm eq}(H\alpha)$.

\subsection{The H$\alpha$ line luminosity}

The method is based on the simple consideration that, at any time, the
largest majority of stars of a given effective temperature $T_{\rm
eff}$ in a stellar population will have no excess H$\alpha$ emission.
Therefore, for stars of that effective temperature, the {\em average}
value of a colour index involving H$\alpha$ (for instance $V-H\alpha$,
and thus not just $R-H\alpha$) effectively defines a spectral reference
template with respect to which the H$\alpha$ colour excess should be
sought. Note that this is not only true for populations comprising
both young and old stars, such as those typical of extragalactic stellar
fields, but also for very young populations since PMS objects show large
variations in their H$\alpha$ emission over hours or days (e.g.
Fernandez et al. 1995; Smith et al. 1999; Alencar et al. 2001), with
only about one third of them at any given time being active H$\alpha$
emitters above a $W_{\rm eq}(H\alpha) \simeq 10$\,\AA\, threshold
(Panagia et al. 2000). 

\begin{figure}
\centering
\resizebox{\hsize}{!}{\includegraphics[width=16cm]{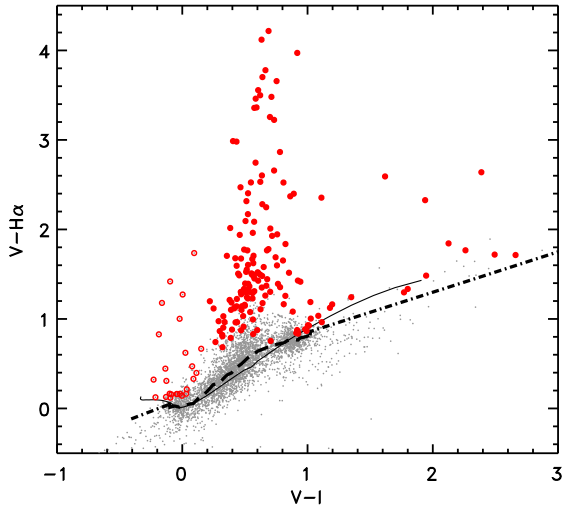}}
\caption{Colour--colour diagram of the selected 4\,156 stars in the field of
SN\,1987A. The dashed line represents the running median  $V-H\alpha$
colour, obtained with a box-car size of 100 points, whereas the thin
solid line shows the model atmospheres of Bessell et al.
(1998). A total of 189 objects with a $V-H\alpha$ excess larger than
$4\,\sigma$ are indicated with large dots.} 
\label{fig4}
\end{figure}

To better clarify the working of our method, we display in
Figure\,\ref{fig4} the $V-H\alpha$ colour as a function of $V-I$ for a
set of stars taken from the catalogue of Romaniello et al. (2002). This
time, we have selected a total of $4\,156$ stars, namely all objects
with $\delta_3 \le 0.1$\,mag, where the mean uncertainty $\delta_3$ is
defined as in Equation\,\ref{eq1}, but for the V, I and H$\alpha$ bands
instead of V, I and R. Therefore, no R-band information is used in this
case and, in fact, no R-band data is needed. It should be noted that
$\delta_3$ is dominated by the uncertainty on the H$\alpha$ magnitude,
while the median value of the uncertainty in the other two bands is
$\delta_V=0.014$ and $\delta_I=0.016$, respectively.  

The thick dashed line in Figure\,\ref{fig4} represents the average
$V-H\alpha$ colour obtained as the running median with a box-car size
of 100 points. The thin solid line shows the colours in these filters
for the model atmospheres of Bessell et al. (1998) mentioned above
(note again the $< 0.1$\,mag discrepancy between models and
observations around $V-I\simeq 0.6$, most likely due to the coarse
spectral sampling of the H$\alpha$ line in the models). At the extremes
of the distribution ($V-I < -0.1$ and $V-I > 1$) the density of
observed points decreases considerably. To determine the reference
template there, we have reduced the box-car size to 10 points and
averaged the running median obtained in this way with the theoretical
models (thin solid line). The thick dot-dashed line in
Figure\,\ref{fig4} shows the best linear fit to the resulting average
(which is extrapolated for $V-I>2$). 

A total of 189 objects, indicated by large dots in Figure\,\ref{fig4},
have a $V-H\alpha$ index exceeding that of the reference template at
the same $V-I$ colour by more than four times the  uncertainty on their
$V-H\alpha$ values. These are the objects with a $V-H\alpha$ excess at
the $4\,\sigma$ level and, as we shall see later, most of them are
bona-fide PMS stars. 

\begin{figure}
\centering
\resizebox{\hsize}{!}{\includegraphics[width=16cm]{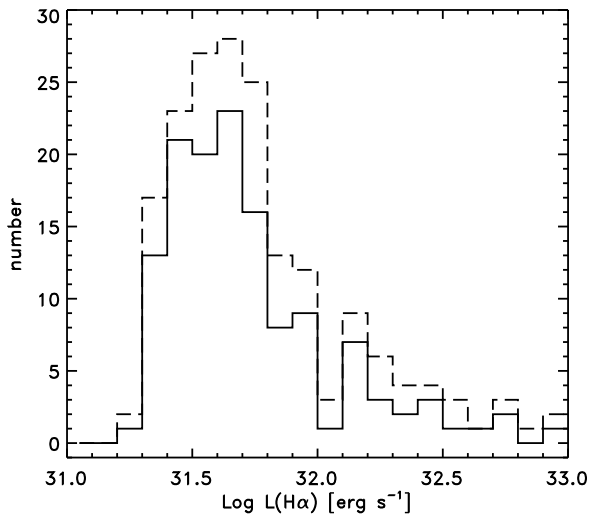}}
\caption{Histogram of the H$\alpha$ luminosities of the 189 stars with
H$\alpha$ excess at the $4\,\sigma$ level (dashed line) and of for the
133 bona-fide PMS that also have an emission $W_{\rm eq}(H\alpha) >
10$\,\AA\,\, and are not Ae/Be stars (solid line). }
\label{fig5}
\end{figure}

Since the contribution of the H$\alpha$ line to the $V$ magnitude is
completely negligible, the magnitude $\Delta H\alpha$ corresponding to
the excess emission is simply:

\begin{equation}
\Delta H\alpha =  (V-H\alpha)^{\rm obs} - (V-H\alpha)^{\rm ref}
\label{eq2}
\end{equation}

\noindent 
where the superscript {\em obs} refers to the observations and {\em
ref} to the reference template. Once $\Delta H\alpha$ is determined in
this way, the H$\alpha$ emission line luminosity $L(H\alpha)$ can be
immediately obtained from the photometric zero point and absolute
sensitivity of the instrumental set-up and from the distance to the
sources. We have assumed a distance to the LMC and, more specifically,
to SN\,1987A of $51.4 \pm 1.2$\,kpc (Panagia et al. 1991, later updated
in Panagia 1999), whereas the photometric properties of the instrument
were taken from the WFPC2 Instrument Handbook (Heyer \& Biretta 2004;
namely inverse sensitivity PHOTFLAM$=1.461 \times 10^{-16}$ and 
photometric zero-point ZP(Vegamag)$=17.564$).  We derive a median value
of the luminosity of the 189 objects with H$\alpha$ excess of $4 \times
10^{31}$\,erg/s or $10^{-2}$\,L$_\odot$. A histogram of the individual
$L(H\alpha)$ values is shown in Figure\,\ref{fig5} (dashed line). 

It should be noted that, because of the width of the H$\alpha$ filter,
$\Delta H\alpha$ includes a small contribution due to [NII] emission
features at $6\,548$\,\AA\, and $6\,584$\,\AA. Generally, their
intensities do not exceed $1.2$\,\% and $3.5$\,\%, respectively, of the
H$\alpha$ line intensity, and in many instances the [NII] doublet lines
are much weaker, i.e. $<0.1\,\%$ of the H$\alpha$ line intensity (e.g.
Edwards et al. 1987; Hartigan, Edwards \& Ghandour 1995). These
intensities should be even lower for PMS stars in the LMC because of the
lower metallicity relative to the Milky Way. However, to take a
conservative approach, we have simply adopted an average value that is
half of the maximum for Galactic T-Tauri stars and we give as
uncertainty half of the measured range. We will therefore adopt average
values of $(1.8 \pm 1.8)$\,\% and $(0.6 \pm 0.6)$\,\% relative to the 
H$\alpha$ intensity, for the 6\,584\,\AA\,\, and 6\,548\,\AA\,\, [NII] 
lines, respectively (note that the uncertainties after the $\pm$ sign 
here are meant to show the entire span of the range, not just the 
$1\,\sigma$ value). Considering the width of the
H$\alpha$ filter of the WFPC2, only the 6\,548\,\AA\,\, [NII] line
contaminates the H$\alpha$ measurements, whereas for measurements made
with the H$\alpha$ filter on board the ACS, both lines will be included
(albeit only in part for the 6\,548\,\AA\,\, line). Although the
resulting corrections on the H$\alpha$ intensity are quite small
(namely $0.994$ for the WFPC2 F656N filter and $0.979$ for the ACS
F658N filter, on average), they are systematic and it is therefore good 
practice to apply them in all cases.

Typically, the combined total uncertainty on our $L(H\alpha)$ 
measurements is $\sim 15\,\%$ and is completely dominated by the
statistical uncertainty on the H$\alpha$ photometry. The systematic
uncertainty on the distance to the field of SN\,1987A accounts for
5\,\% (see Panagia 1999), whereas  the typical uncertainty on the
absolute sensitivity of the instrumental setup is of order 3\,\%.

\subsection{The H$\alpha$ equivalent width}

If the stars defining the reference template had no H$\alpha$
absorption features (we have established above that they do not have
H$\alpha$ in emission), their $V-H\alpha$ index would correspond to
that of the pure continuum and could then be used to derive the
equivalent width of the H$\alpha$ emission line, $W_{\rm
eq}(H\alpha)$. While this approximation might be valid for stars with
a conspicuous H$\alpha$ emission, it is clearly not applicable in general.
It is, however, possible in all cases to derive the level of the
H$\alpha$ continuum by using properly validated models atmospheres.

As discussed above, the models of Bessell et al. (1998) reproduce quite
reliably the observed broad-band colours (see e.g. Figure\,\ref{fig2}).
This means that, even though the models might lack the resolution
necessary to realistically reproduce spectral lines (hence the small
discrepancy between the dashed line and the squares in
Figure\,\ref{fig3}), the level of the continuum can be trusted. One
can, therefore, fit the continuum in those models in a range containing
the H$\alpha$ line (e.g. $6\,500 - 6\,620$\,\AA\,) and fold the
resulting ``H$\alpha$-less'' model spectra through the instrumental
set-up (e.g. with Synphot), so as to derive the magnitude of the sole
continuum in the H$\alpha$ band ($H\alpha^c$) as a function of spectral
type or effective temperature. We present in the Appendix the
relationships between $V$, $I$ and $H\alpha^c$ for several HST
instruments and filters. The difference between the observed $H\alpha$
magnitude and $H\alpha^c$ provides a direct measure of $W_{\rm
eq}(H\alpha)$, as we explain below.

We recall that the equivalent width of a line is defined as:

\begin{equation}
W_{\rm eq}= \int{(1 - P_\lambda) \, d\lambda} 
\label{eq3}
\end{equation}

\noindent  
where $P_\lambda$ is the profile of the line, or the spectrum of the 
source divided by the intensity of the continuum, and the integration
is extended over the spectral region corresponding to the line. If the
line profile is very narrow when compared to the width of the filter,
i.e. if the line falls completely within the filter bandpass as is
typical of emission lines, then $W_{\rm eq}(H\alpha)$ is simply given
by the relationship:

\begin{equation}
W_{\rm eq} (H\alpha) =  \mathrm{RW} \times [1-10^{-0.4 \times
(H\alpha-H\alpha^c)}]
\label{eq4}
\end{equation}

\noindent   
where $\mathrm{RW}$ is the rectangular width of the filter (similar in
definition to the equivalent width of a line), which depends on the
characteristics of the filter. Values of $\mathrm{RW}$ for the
H$\alpha$ bands of past and current HST instruments are given in the
Appendix (Table\,\ref{tab4}). As in the case of $L(H\alpha)$, the
statistical uncertainty on the equivalent width is dominated by the
uncertainty on the $H\alpha$ magnitude, typically of order 15\,\% in
our selection. Moreover, the value of $W_{\rm eq}(H\alpha)$ obtained in
this way is also subject to some systematic uncertainties produced by
the model atmospheres used to interpolate the level of the continuum,
such as e.g. metallicity, surface gravity or spectral resolution, and
should therefore be taken with caution. On the other hand, we show in
the Appendix that the equivalent widths (in absorption) obtained by
applying this method to the high-resolution model spectra of Munari et
al. (2005) are in excellent agreement with those obtained via standard
spectro-photometry (see Figure\,\ref{fig18}).

\begin{figure}[t]
\centering
\resizebox{\hsize}{!}{\includegraphics[width=16cm]{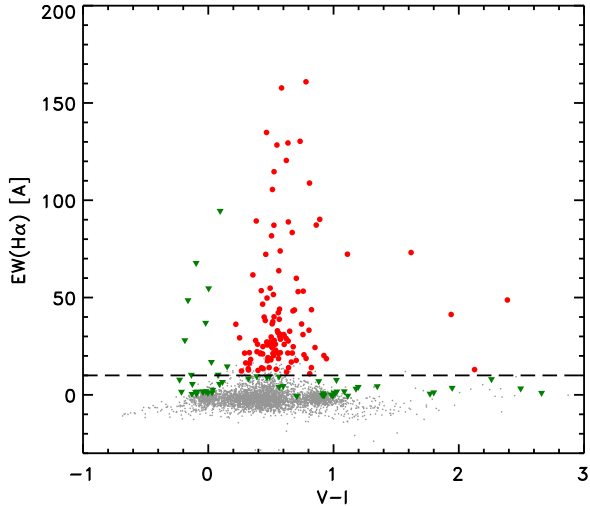}}
\caption{Equivalent width of the selected 4\,156 stars in the field of
SN\,1987A, as a function of their $V-I$ colour. Of the 189 stars with
H$\alpha$ excess at the $4\,\sigma$ level (circles and triangles), a
total of 164  have $W_{\rm eq}(H\alpha) > 10$\,\AA. While objects bluer
than $V-I \simeq 0.2$ are probably Ae/Be stars, there are 133 redder
stars with significant H$\alpha$ excess (marked as circles) that we
consider bona-fide low-mass PMS stars.} 
\label{fig6}
\end{figure}

Similarly, the values of $W_{\rm eq}(H\alpha)$ that we obtain in the
field of SN\,1987A are fully consistent with those of young PMS stars.
In Figure\,\ref{fig6} we show the value of $W_{\rm eq}(H\alpha)$  as a
function of $V-I$ for the $4\,156$ selected stars. As mentioned
earlier, there are 189 objects with a $\Delta H\alpha$ excess above the
$4\,\sigma$ level and they are indicated as thick symbols (circles and
triangles) in Figure\,\ref{fig6}. Since the equivalent width shown in
Figure\,\ref{fig6} is that of the pure emission component, the spectra
of stars with small $W_{\rm eq}$, say $\la 10$\,\AA, actually show an
H$\alpha$ in absorption. For this reason, we conservatively ignore
stars with measured $W_{\rm eq}(H\alpha)<10$\,\AA, since this value is
about the largest absorption equivalent width expected for normal stars
(see Figure\,\ref{fig18} in the Appendix). A total of 164 objects
satisfy this condition, but, in light of our conservative threshold,
this must be considered as a lower-limit to the number of objects with
genuine H$\alpha$ excess.  

Values of $W_{\rm eq}(H\alpha)$ for the sample range from 10\,\AA\, to
650\,\AA, with a median of 19\,\AA. These values are in excellent
agreement with those obtained by Panagia et al. (2000) from the
$R-H\alpha$ colour excess of the stars in this field and represent a
refinement of those measurements, since these are based on a more
accurate determination of the level of the continuum. These equivalent
widths are  typical of PMS stars (e.g. Muzerolle, Hartmann \& Calvet
1998) and about two orders of magnitude smaller than those expected for
pure nebular emission (see discussion in Section\,6). While objects
bluer than $V-I \simeq 0.2$ are probably Ae/Be stars, there are 133
redder stars with significant H$\alpha$ excess (marked as circles in
Figure\,\ref{fig6}) that are most likely bona-fide low-mass PMS stars
of various ages, as we will show in the following sections. A histogram
of their H$\alpha$ luminosities is shown in Figure\,\ref{fig5} (solid
line).

\section{Accretion luminosity and mass accretion rate}

It is generally accepted (see e.g. Hartmann et al. 1998) that the
energy $L_{\rm acc}$ released by the magnetospheric accretion process
goes towards ionising and heating the circumstellar gas. In this
hypothesis, the bolometric accretion luminosity $L_{\rm acc}$ can be
determined from a measurement of the {\em reradiated energy} and, in
particular, via the measured H$\alpha$ luminosity that is produced in
the process.

In order to determine the exact relationship between $L(H\alpha)$ and
$L_{\rm acc}$, we have used literature values of $L_{\rm acc}$ and
$L(H\alpha)$ measurements as recently summarized by Dahm (2008) for PMS
stars in the Taurus--Auriga association. As pointed out by Dahm 
himself, there may be a serious problem in determining an empirical
relationship between $L_{\rm acc}$ and $L( H\alpha)$ measurements
because the two quantities were determined from non-simultaneous
observations. The intensity of the H$\alpha$ line from PMS sequence
stars is known to vary by about 20\,\% on a time  scale of a few hours
and by as much as a factor of 2 -- 3 in a few days (e.g.  Fernandez
et al. 1995; Smith et al. 1999; Alencar et al. 2001). For this reason, 
a relationship between $L(H\alpha)$ and $L_{\rm
acc}$ is necessarily rather uncertain, although a very clear trend can
be seen in the  $L_{\rm acc}~vs.~L(H\alpha)$ plot shown by Dahm (2008).
His logarithmic best fit would provide a slope of $1.18 \pm 0.26$ for
such a relationship. On the other hand, theoretical models (e.g.
Muzerolle, Calvet \& Hartmann 1998) would predict logarithmic slopes
of  about unity for low accretion rates, hence for faint H$\alpha$
luminosities, and shallower slopes for higher luminosities. 

In the absence of compelling evidence in favour or against a steep
slope, we will adopt a logarithmic slope of unity for the empirical
$L_{\rm acc}~vs.~L(H\alpha)$ relationship, i.e. a constant ratio
$L_{\rm acc} / L(H\alpha)$,  and determine the proportionality constant
from an elementary fit of the data summarised by Dahm (2008). On this 
basis we obtain:

\begin{equation}
\log L_{\rm acc} = (1.72 \pm 0.47) + \log L(H\alpha)
\label{eq5}
\end{equation}

As anticipated, the resulting uncertainty on $L_{\rm acc}$ is rather
large, as high as a factor of 3, but this is the best one can obtain
with the available data sets. In view of the convenience and the
feasibility of determining $L_{\rm acc}$ from $L(H\alpha)$ for PMS
outside our own Galaxy, it would be of paramount importance to obtain a
proper calibration of the $L_{\rm acc}~vs.~L(H\alpha)$ relationshipby
making simultaneous spectral observations of PMS stars over a suitably
wide wavelength baseline, i.e. covering at least the range
3000--8000\,\AA, and possibly extending it to the near IR with
measurements of Paschen and Brackett lines. High quality data of this
type are already available for the Orion Nebula Cluster, collected by
dedicated projects both from ground based observatories and with the
HST (i.e. the Legacy Program on Orion; P.I. Massimo Robberto) and their
analysis is in progress to provide a more accurate calibration of the
$L_{\rm acc}~vs.~L(H\alpha)$ relationship (Da Rio et al., in
preparation; Robberto et al., in preparation; Panagia et al., in
preparation).

The median value of the accretion luminosity thus obtained  for our
sample of 133 PMS stars is $\sim 0.57$\,L$_\odot$, in good agreement
with the value of $0.54$\,L$_\odot$ derived from the U-band excess
measured by Romaniello et al. (2004) for stars in the same field,
especially considering that their selection of stars with excess
emission is less stringent than our $4\,\sigma$ level (we discuss this
point in Section\,7).


Once $L_{\rm acc}$ is known, the mass accretion rate follows from the
free-fall equation that links the luminosity released in the impact of
the accretion flow with the rate of mass accretion $\dot M_{\rm acc}$
according to the relationship:

\begin{equation}
L_{\rm acc} \simeq \frac{G\,M_*\,\dot M_{\rm acc}}{R_*} \left(1 - 
\frac{R_*}{R_{\rm in}}\right)
\label{eq6}
\end{equation}

\noindent
where $G$ is the gravitational constant, $M_*$ and $R_*$ the mass and
photospheric radius of the star and $R_{\rm in}$ the inner radius of
the accretion disc. The value of $R_{\rm in}$ is rather uncertain and
depends on how exactly the accretion disc is coupled with the magnetic
field of the star. Following Gullbring et al. (1998), we adopt $R_{\rm
in} = 5\,R_*$ for all PMS objects. 

As regards $R_*$, we derive it from the luminosity and effective
temperature of the stars, which in turn come from the observed colour
$(V-I)$ and magnitude ($V$), properly corrected for interstellar
extinction as provided by Romaniello (1998), Panagia et al. (2000) and
Romaniello et al. (2002). The adopted distance to SN\,1987A is
$51.4$\,kpc (Panagia et al. 1991; Panagia 1999), as mentioned above.
The stellar mass $M_*$ was estimated by comparing the location of a
star in the Hertzsprung--Russell (H--R) diagram with the PMS 
evolutionary
tracks. As for the latter, we adopted those of Degl'Innocenti et al.
(2008; see also Tognelli, Prada Moroni \& Degl'Innocenti, in
preparation), for metallicity $Z=0.007$ or about one third $Z_\odot$, as
appropriate for the LMC (e.g. Hill, Andrievsky, \& Spite 1995; Geha et
al. 1998). These new PMS tracks were specifically computed with an
updated version of the FRANEC evolutionary code (see Chieffi \&
Straniero 1989 and Degl'Innocenti et al. 2008 for details and Cignoni et
al. 2009 for an application). However, in order to assess how
differences in the evolutionary models affect our results, in
Section\,6.1 we also consider PMS tracks from other authors, namely
those of D'Antona \& Mazzitelli (1997) and Siess, Dufour \& Forestini
(2000).

\begin{figure}
\centering
\resizebox{\hsize}{!}{\includegraphics[width=16cm]{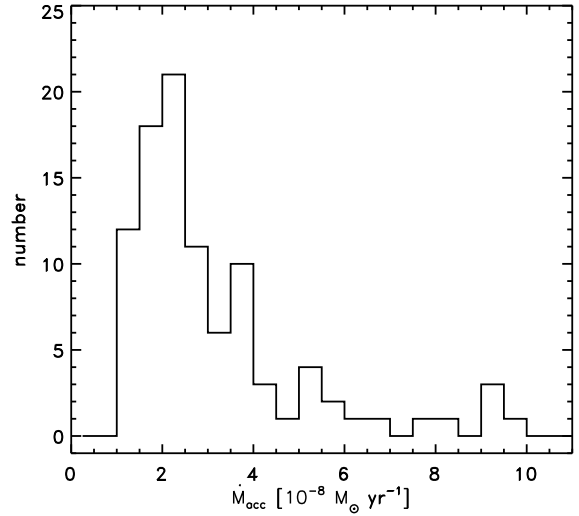}}
\caption{Distribution of the mass accretion rates of the 133
bona-fide PMS stars. Note that 9 objects with $\dot M_{\rm acc} >
10^{-7}$\,\Msolar\,yr$^{-1}$ are not shown in the graph.} 
\label{fig7}
\end{figure}

Combining Equations\,\ref{eq5} and \ref{eq6}, we can link the mass
accretion rate $\dot M_{\rm acc}$, in units of \Msolar yr$^{-1}$, to
$L(H\alpha)$:

\begin{eqnarray}
\log \frac{\dot M_{\rm acc}}{M_\odot {\rm yr}^{-1}} & = &-7.39 + \log
\frac{L_{\rm acc}}{L_\odot} + \log\frac{R_*}{R_\odot} -
\log\frac{M_*}{M_\odot} \\
&  & \nonumber \\
& = & (-5.67 \pm 0.47) + \log\frac{L(H\alpha)}{L_\odot} + \log
\frac{R_*}{R_\odot} - \log \frac{M_*}{M_\odot} \nonumber
\label{eq7}
\end{eqnarray}

The distribution of mass accretion rates found in this way for the 133
PMS stars in our sample is shown in Figure\,\ref{fig7}. The median
value of $2.9 \times 10^{-8}$\,\Msolar\,yr$^{-1}$ is in remarkably good
agreement with the median rate of $2.6 \times
10^{-8}$\,\Msolar\,yr$^{-1}$ found by Romaniello et al. (2004) from the
U-band excess.


The uncertainty on $\dot M_{\rm acc}$ is dominated by the approximate
knowledge of the ratio of $L_{\rm acc}$ and $L(H\alpha)$ in
Equation\,\ref{eq5}. As explained above, empirical measurements and
theoretical models suggest a value of $L_{\rm acc}/L(H\alpha) \simeq
52$ but with an uncertainty of a factor of three. Since the ratio,
however, is the same for all stars, the comparison between different
objects is not hampered by this uncertainty, as long as the statistical
errors are small. 

As for the other quantities in Equation\,7, we discuss here  the
sources of statistical uncertainty, while systematic errors are
addressed separately in Section\,6. With our selection criteria, the
typical uncertainty on $L(H\alpha)$ is 15\,\% and is dominated by
random errors. The uncertainty on $R_*$ is typically 7\,\%, including a
5\,\% systematic uncertainty on the distance modulus. As for the mass
$M_*$, since it is determined by comparing the location of a star in
the H--R diagram with evolutionary tracks, both systematic and
statistical uncertainties are important. The uncertainty on the
temperature is mostly statistical and of order 3\,\%, while that on the
luminosity is $7\,\%$, comprising both random errors (1\,\% uncertainty
on the bolometric correction and 3\,\% on the photometry) and
systematic effects (5\,\% uncertainty on the distance modulus). When we
interpolate through the PMS evolutionary tracks to
estimate the mass, the uncertainties on $T_{\rm eff}$ and $L$ imply an
error of $\sim 7\,\%$ on $M_*$. However, an even larger source of
systematic uncertainty on $M_*$ comes from the evolutionary tracks, as
we explain in Section\,6. In summary, the combined statistical
uncertainty on $\dot M_{\rm acc}$ is 17\,\%.

\begin{figure}
\centering
\resizebox{\hsize}{!}{\includegraphics[width=16cm]{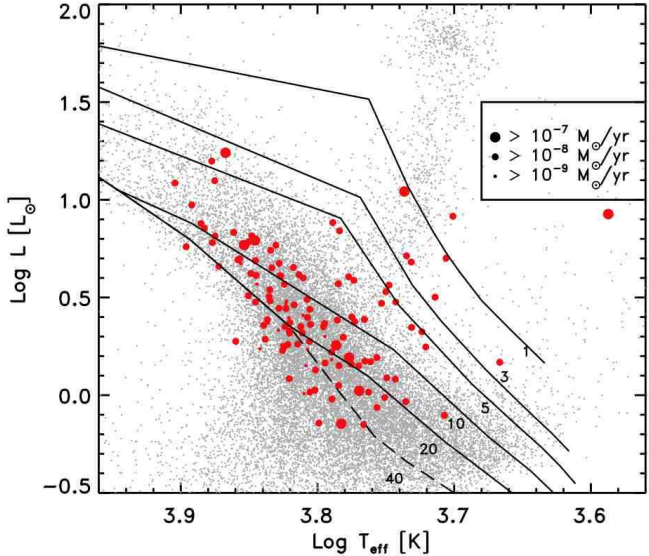}}
\caption{Location of the bona-fide PMS candidates in the H--R diagram,
compared with the isochrones of Degl'Innocenti et al. (2008) for
metallicity $Z = 0.007$ (the ages, in Myr, are indicated next to each
isochrone). The size of the symbols is proportional to the mass
accretion rate $\dot M_{\rm acc}$ (see legend).} 
\label{fig8}
\end{figure}

\section{Ages of PMS stars}

Having identified a population of bona-fide PMS stars, with well
defined accretion luminosities and mass accretion rates, it is
instructive to place them in the H--R diagram. We do so in
Figure\,\ref{fig8} where bona-fide PMS stars are marked as large dots,
with a size proportional to their $\dot M_{\rm acc}$ value, according
to the legend. Also shown in the figure are the theoretical isochrones
from the FRANEC models
of Degl'Innocenti et al. (2008) for $Z=0.007$ and ages as indicated (in
units of Myr, from the stellar birth line; see Palla \& Stahler 1993).
The dashed line, corresponding to a 40 Myr isochrone, defines in
practice the zero-age main sequence (MS). 

It is noteworthy that the majority of PMS objects in Figure\,\ref{fig8}
are rather close to the MS and would have easily been missed if no
information on their H$\alpha$ excess had been available. Indeed, it is
customary to identify PMS stars in a colour--magnitude diagram by
searching for objects located above and to the right of the MS, since
this is where one would expect to find very young objects. This method
was first used by Gilmozzi et al. (1994) to identify a population of
PMS stars outside the Milky Way, namely in the LMC cluster NGC\,1850.
More recent applications of this method include e.g. those of Sirianni
et al. (2000), Nota et al. (2006) and Gouliermis et al. (2006). 

Unfortunately, this method of identification is not very reliable,
since the existence of an age spread and the presence in the same field
of a considerably older population, as well as unaccounted patchy
absorption, all tend to fill up the parameter space between the MS and
the birth line, thereby thwarting any attempts to identify PMS stars on the
basis of their effective temperature and luminosity alone. In the
specific case of the SN\,1987A field studied here, Panagia et al.
(2000) have shown that the age spread is remarkable, with several
generations of young stars with ages between 1 and 150\,Myr superposed
on a much older field population ($0.6-6$\,Gyr). In such circumstances,
as Figure\,\ref{fig8} shows, broad-band photometry alone would have
not identified as such most of our bona-fide PMS stars.

\begin{figure}
\centering
\resizebox{\hsize}{!}{\includegraphics[width=16cm]{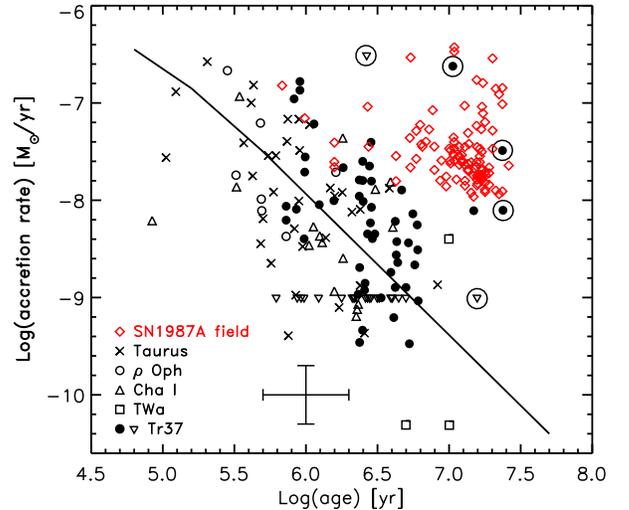}}
\caption{The mass accretion rate as a function of stellar age for our
bona-fide PMS stars (diamonds) is compared here with that of a similar
number of Galactic T Tauri stars (see legend) from the work of
Sicilia--Aguilar et al. (2006), from whom all Galactic data-points are
taken (the large cross indicates the uncertainties as quoted in that
paper). The solid line shows the evolution for current models of viscous
disc evolution from Hartmann et al. (1998). Our measurements are
systematically higher than the models, but are in agreement with the
$\dot M_{\rm acc}$ values of G type stars in the Galaxy (large circles).} 
\label{fig9}
\end{figure}

In accordance with their proximity to the MS, the majority of bona-fide
PMS stars turn out to be relatively old. The ages of individual objects
were determined by interpolating between the isochrones in the H--R 
diagram. They
range from 1 to $\sim 50$\,Myr, with  a median age of $13.5$\,Myr, 
in very good agreement with the estimated age of both SN\,1987A and its 
nearby Star 2 of $\le 13$\,Myr (see Scuderi et al. 1996 and references 
therein). Panagia et al. (2000) find that about 35\,\% of the stars
younger than 100\,Myr in this field have a typical age of 12\,Myr,
again consistent with the estimated age of the bona-fide PMS stars. 

The relatively old age of our PMS stars might appear at odds with the
rather large mass accretion rates that we obtain for that population,
namely $\sim 10^{-8}$\,\Msolar\,yr$^{-1}$. According to current models
of viscous disc evolution (see Hartmann et al. 1998; Muzerolle et al.
2000), the mass accretion rate of a $\sim 14$\,Myr old PMS population
should be of order $3 \times 10^{-10}$\,\Msolar\,yr$^{-1}$. Those
models, shown in Figure\,\ref{fig9} as a solid line, adequately
reproduce the trend of decreasing $\dot M_{\rm acc}$ with stellar age
observed for very low-mass Galactic T-Tauri stars (see e.g. Calvet et
al. 2004), but under-predict our measurements (indicated by diamonds) by
over an order of magnitude.

On the other hand, recent observations of PMS stars in the cluster
Trumpler\,37 by Sicilia--Aguilar et al. (2006), from whom the
data-points in Figure\,\ref{fig9} are taken, indicate that the value of
$\dot M_{\rm acc}$ for $\sim 10-20$\,Myr old G-type PMS stars (large
circles) is considerably larger, indeed of order $\sim
10^{-8}$\,\Msolar\,yr, in excellent agreement with our results. With a
median mass of $1.2$\,\Msolar and a median  $V-I$ colour of $0.56$, our
bona-fide PMS stars are fully consistent with the G spectral type and
suggest that stars of higher mass have a higher $\dot M_{\rm acc}$
value at all ages. As mentioned in the Introduction,  Muzerolle et al.
(2003; 2005), Natta et al. (2004; 2006) and Calvet et al. (2004)
suggest a dependence of the type $\dot M_{\rm acc} \propto M^2$ from
the analysis of a sample of low-mass and intermediate-mass T-Tauri
stars, albeit with a large uncertainty on the index (see also Clarke \&
Pringle 2006). A similar trend is proposed by Sicilia--Aguilar et al.
(2006), yet with even lower statistical significance due to the limited
mass range covered by their observations. The range of masses that we
cover here is also quite limited and does not allow us to address this 
issue in detail. We will, however, do so in a forthcoming paper (De
Marchi et al. 2008, in preparation), where we compare the accretion
rates in this field with those of a much younger and an order of
magnitude more numerous population of PMS stars in the Small Magellanic
Cloud. While in the SN\,1987A field the detection limits imposed by the WFPC2
photometry do not allow us to reach stars with mass below  $\sim
0.9$\,M$_\odot$, in that paper, based on ACS observations, we will study
stars down to $\sim 0.4$\,M$_\odot$. 

\begin{figure}
\centering
\resizebox{\hsize}{!}{\includegraphics[width=16cm]{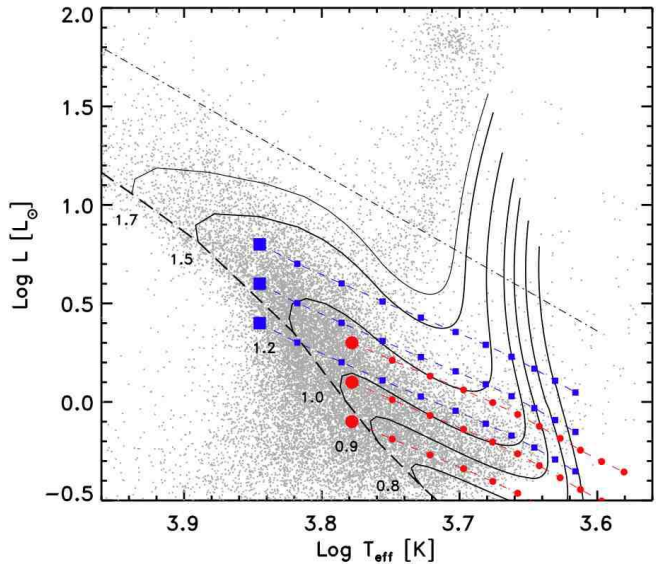}}
\caption{Simulated effects of a reddening increase on the mass
determination via interpolation through the evolutionary tracks. The
solid lines are the track of Siess et al. (2000) corresponding to the
masses as indicated. Small circles and squares show the reddening 
displacement, from the unreddened locations (larger circles and
squares),  in increments of $E(V-I)=0.1$ according to the extinction
law of Scuderi et al. (1996) for this field. The dot-dashed line
indicates the line of constant radius, while the thick dashed line is
the same 40 Myr isochrone as in Figure\,\ref{fig8}.} 
\label{fig10}
\end{figure}

\section{Systematic errors}

The main sources of systematic uncertainty on the derived mass
accretion rates are ({\em i}) discrepancies in the isochrones and
evolutionary tracks, ({\em ii}) reddening, ({\em iii}) H$\alpha$
emission generated by sources other than the accretion process, and
({\em iv}) the contribution of the nebular continuum to the colours of
the stars. We discuss all these effects in this section.

\subsection{Models of stellar evolution}

As explained in Sections\,4 and 5, the mass and age of PMS stars are
determined by interpolation of their location in the H--R diagram,
using model evolutionary tracks and isochrones as references. Apart for
possible inaccuracies in the models input physics and errors in the
interpolation, the largest source of systematic uncertainty on the
derived mass and age comes from the use of models that might not
properly describe the stellar population under study (e.g. because of
the wrong metallicity) and from differences between models of various
authors.

For instance, if we had used Siess et al.'s (2000) tracks for
metallicity 1/2 solar in place of those for 1/3 solar, the masses of
our PMS objects would be systematically higher by about 20\,\%.
For metallicity 1/2 solar, using the tracks of Siess
et al. (2000) or those of D'Antona \& Mazzitelli (1997) results in
masses that agree to within a few percent. On the other hand, when it 
comes to the determination of the age, using D'Antona \& Mazzitelli's 
(1997) isochrones one would derive ages twice as young for stars colder
than $\sim 6\,500$\,K, while for hotter stars ages would only be 20\,\%
lower.

In summary, systematic uncertainties of order 20\,\% are to be expected
for the mass and possibly higher for the age determination.

\subsection{Reddening}

Dust extinction systematically displaces stars in the H--R diagram
towards lower luminosities and effective temperatures. This is shown
graphically in Figure\ref{fig10}, for six model stars, three with
$T_{\rm eff}=6\,000$\,K (large circles) and three with $T_{\rm
eff}=7\,000$\,K (large squares). The effects of a reddening increase
from $E(V-I)=0$ to $E(V-I)=1$ on the location of the stars in the H--R 
diagram is shown by the smaller symbols, corresponding to increments 
of $E(V-I)=0.1$. We have adopted here the extinction law as measured
specifically in the field of SN\,1987A by Scuderi et al. (1996).

Since the reddening vector in the H--R plane is almost parallel to the
lines of constant radius (as shown by the dot-dashed line for
$R=10$\,R$_\odot$), extinction affects only marginally the estimate of
the stellar radius. However, at low temperatures the reddening vector
crosses the evolutionary tracks (solid lines), thus leading to an
underestimate of the stellar mass, if no extinction correction is
applied. In turn, this will result in a systematically higher value of
the $R_*/M_*$ ratio and, therefore, in an overestimate of the mass
accretion rate derived via Equation\,\ref{eq6}.

\begin{figure}
\centering
\resizebox{\hsize}{!}{\includegraphics[width=16cm]{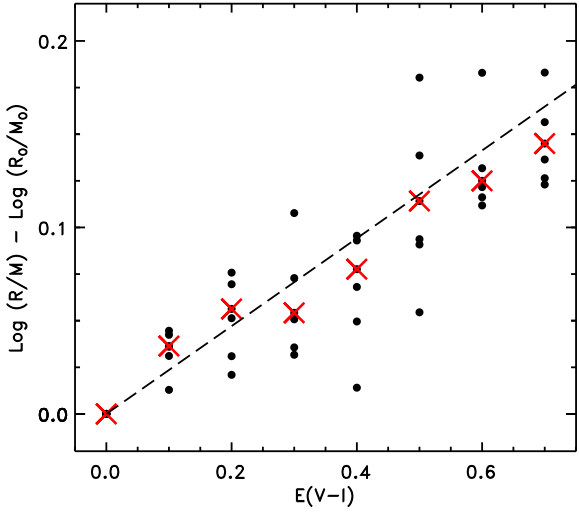}}
\caption{Overestimate of the $R/M$ ratio as a function of $E(V-I)$ when
the extinction remains unaccounted. Thick dots represent the $R/M$
ratios of the artificially reddened stars (see Figure\,\ref{fig10}) and
the crosses their median values.} 
\label{fig11}
\end{figure}

In order to better gauge the uncertainty on the value of $R_*/M_*$
caused by unaccounted extinction, we have determined the mass and
radius of each one of the artificially reddened model stars shown in
Figure\,\ref{fig10}, in exactly the same way in which we did that for
bona-fide PMS stars, and compared those with the known input values.
The result is shown graphically in Figure\,\ref{fig11}, where, the
relation between the derived and input $R_*/M_*$ ratios is shown as a
function of colour excess $E(V-I)$. 

Regardless of the input parameters ($L$, $T_{\rm eff}$ or $R_*/M_*$),
it appears that over the range explored here the derived $R_*/M_*$ 
ratio increases slowly with colour excess (see dashed line that best
fits the median values indicated by the crosses). In particular,
underestimating the $E(V-I)$ colour  excess by $\sim 0.4$\,mag would
lead to a 30\,\% overestimate of $R_*/M_*$ (and hence of $\dot M_{\rm
acc}$). This appears unlikely, considering that the best estimate of
the total extinction --- including the Milky Way component --- towards
Star 2, one of the two bright close companions to SN\,1987A, is
$E(V-I)=0.21$ (Scuderi et al. 1996), with a $1\,\sigma$ dispersion over
the field of $\sim 0.07$ (Romaniello 1998; Panagia et al. 2000).
According to Zaritsky et al. (2004), the average reddening towards cool
(5500\,K$< T_{\rm eff}< $6\,500\,K) stars in the LMC is $A_{\rm V} =
0.43$, thus implying $E(V-I) \simeq E(B-V) \simeq 0.14$. Therefore, for
an average LMC star forming region, omitting the extinction correction
would result in a 10\,\% overestimate of the mass accretion rate. This
error, albeit systematic, is smaller than most  other systematic
uncertainties and comparable to the measurement errors.

Nonetheless, we would like to stress here that the effects of patchy
absorption can be more severe in regions of high extinction. Therefore,
whenever possible, one should apply extinction corrections to each
individual stars, as it was done for the photometric catalogue of
Romaniello et al. (2002) that we use here.

\vspace*{0.5cm}

\subsection{H$\alpha$ emission not due to the accretion process}

The underlying assumption in the determination of the mass accretion
rate of PMS stars is that the gravitational energy $L_{\rm acc}$
released in the accretion process goes into heating of the gas at the
boundary layer (see e.g. Hartmann et al. 1998). In this case, the
luminosity of the H$\alpha$ emission line can provide a measure of the
accretion energy, because it acts as a natural ``detector'' of the
luminosity released in the accretion process.

Thus, the issue to address is whether the measured $L(H\alpha)$ is
produced only and entirely as a result of the energy released in the 
accretion process. Obviously, depending on the geometry and orientation
of the circumstellar disc and on the specifics of the accretion
process, some of the ionising energy could escape without effectively
ionising the local gas. This would result in an underestimate of the
true mass accretion rate for some stars, but this is true of any method
that uses H emission lines as indicators of the accretion process and,
therefore, it does not affect the comparison of our results with those
of others (see Figure\,\ref{fig9}). 

One possibility that we can readily exclude is that the H$\alpha$
emission that we detect is due to chromospheric activity. The total
chromospheric emission of a solar-type star on the subgiant branch
corresponds to a luminosity of about $10^{-4}$\,L$_{\odot}$
(Ulmschneider 1979; Pasquini et al. 2000), i.e. two orders of magnitude
less than what we measure.

On the other hand, we must also consider the possibility that
H$\alpha$  emission might occur in discrete H knots along the line of
sight, in places unrelated with the PMS object. Furthermore, hot nearby
massive stars may ionise the H gas in which PMS stars are immersed,
eventually producing H$\alpha$ emission when the gas recombines. Both
these effects would result in a overestimate of $\dot M_{\rm acc}$, but
we show below that the probability of this happening is very small.

As for intervening H$\alpha$ emission along the line of sight, it would
have to arise in knots of ionised H, e.g. very compact HII regions,
contained within our photometric apertures. However, if such knots
exist, they should not project only against stars and, therefore,
narrow-band images should reveal H$\alpha$ emission where broad-band
imaging shows no or very low signal (see Section 6.3). Since this
is not observed, statistical arguments rule out this possibility as a
significant source of the observed H$\alpha$ excess emission.

While it is possible that some of the H$\alpha$ emission that we detect
is due to diffuse nebular emission in the HII region not powered by the
accretion process, if the emission is extended and uniform over an area
of at least $0\farcs5$ radius around the star its contribution cancels
out with the rest of the background when we perform aperture photometry
(using an aperture of $0\farcs2$ radius for the object and an annulus
of $0\farcs3 - 0\farcs5$ for the background). The subtraction would not
work if the emission were not uniform, as for example in the case of a
filament that projects over the star but that does not cover completely
the background annulus. However, we have visually inspected all objects
with excess H$\alpha$ emission and excluded the few cases where a
spurious filament of this type could be seen (see Romaniello 1998;
Panagia et al. 2000). 

Another case in which the emission does not cancel out is when the gas
ionised by an external source is associated to the star and completely
contained within our photometric aperture. We recall here that the
typical size of a circumstellar disc is of order 100 AU (e.g. Hartmann
et al. 1998), or about $0.02$ pixel at the distance of SN\,1987A, and
that the median $L(H\alpha)$ that we measure for our PMS stars is $4
\times 10^{31}$\,erg\,s$^{-1}$, or about $10^{-2}$\,L$_\odot$ (see
Figure\,\ref{fig5}). We can then calculate  how close a PMS stars
should be to a hot, UV-bright star for the latter to flood the PMS
circumstellar disc with enough ionising radiation to cause at least
10\,\% of the H$\alpha$ luminosity that we measure from that source.

Let $r=100$\,AU be the external radius of the circumstellar disc and
$d$ the distance to an ionising star that illuminates the disc, assumed
to have zero inclination (i.e. face on), so that $\phi=r^2/(4\,d^2)$ is
the solid angle subtended by the disc. Let $\beta_2$ be the
recombination coefficient to the first excited level in an optically
thick H gas. If $V$ is the volume of the gas and $n_e$ and $n_p$ are
the electron and proton densities per unit volume, the global
ionisation equilibrium equation (case B; see Baker \& Menzel 1938)
is then:

\begin{equation}
N_{L} \, \phi = \beta_2 \, n_e \, n_p \, V
\label{eq8}
\end{equation}

\noindent 
where $N_L$ is the rate of ionising photons produced by the external
star. Denoting with $\alpha_{3,2}$ the H$\alpha$ emission
coefficient for the same gas, the luminosity of the H$\alpha$
recombination line can be written as:

\begin{equation}
L(H\alpha) = \alpha_{3,2} \, n_e \, n_p \, V
\label{eq9}
\end{equation}

\noindent 
which allows us to express the luminosity of the H$\alpha$ emission
caused by the external star as a function of its ionising photon rate
$N_L$:

\begin{equation}
L^{\rm ext}(H\alpha) = \alpha_{3,2} / \beta_2 \, N_L \,\phi .
\label{eq10}
\end{equation}

The minimum distance beyond which the contribution of the ionising star to
the total H$\alpha$ emission becomes negligible (i.e. 10\,\% or less) is
given by:

\begin{equation}
d^2_{\rm min} = \frac{r^2}{4} \, \frac{\alpha_{3,2}}{\beta_2} \, \frac{10
\, N_L}{L(H\alpha)} .  
\label{eq11}
\end{equation}

Table\,\ref{tab1} gives the value of $\phi$ for a typical H$\alpha$
luminosity of $10^{-2}$\,L$_\odot$, as a function of the spectral type
of the ionising star. Here we have used the compilation of Panagia
(1973) for the appropriate stellar parameters. The table also provides
the minimum distance $d_{\rm min}$ in pc and in arcsec (at the distance
of SN\,1987A) for a circumstellar disc of radius $r=100$\,AU. We recall
here that 1 arcsec corresponds to 10 WFPC2 pixels. The brightest star
in the field studied here is an object with $\log T_{\rm eff} = 4.7$
and $\log (L/L_\odot) = 5.8$, with an implied mass of $\sim
60$\,\Msolar and an age less than 1\,Myr (see Panagia et al. 2000). The
corresponding value of $N_L$, from  Panagia (1973), is $4 \times
10^{49}$\,s$^{-1}$, thus implying $d_{\rm min}=0.95$\,pc or 38 WFPC2
pixel for a gas temperature of 10\,000\,K. A careful inspection of the
images shows that none of our candidate bona-fide PMS stars falls
closer to this star than $d_{\rm min}$. The same applies to the second
and third brightest stars in the field, respectively with $\log T_{\rm
eff} = 4.7$, $\log (L/L_\odot) = 5$ and $\log T_{\rm eff} = 4.67$,
$\log (L/L_\odot) = 4.75$, as well as to all fainter objects.

\begin{deluxetable}{lcccccc} 
\tablecolumns{7}
\tablewidth{0pc}
\tablecaption{Possible effects of nearby ionising stars}
\tablehead{
\colhead{SpT} & \colhead{$T_{\rm eff}$} & 
\colhead{$\log(L/L_\odot)$} & \colhead{$\log N_L$} & 
\colhead{$\phi$} & \multicolumn{2}{c}{$d_{\rm min}$} \\
\colhead{}& \colhead{(K)} & \colhead{} & \colhead{(s$^{-1}$)} & 
\colhead{} & \colhead{(pc)} & \colhead{($\arcsec$)} } 
\startdata
O4 & 50\,000 & 6.11 & 49.93 & 3.42e-08 & 1.31 & 5.26\\
O6 & 42\,000 & 5.40 & 49.08 & 2.43e-07 & 0.49 & 1.97\\
O7 & 36\,500 & 4.81 & 48.62 & 6.93e-07 & 0.29 & 1.17\\
O9 & 34\,500 & 4.66 & 48.08 & 2.43e-06 & 0.16 & 0.62\\
B0 & 30\,900 & 4.40 & 47.36 & 1.27e-05 & 0.07 & 0.27\\
B1 & 22\,600 & 3.72 & 45.29 & 1.50e-03 & 0.01 & 0.03\\
\enddata
\tablecomments{The values of $\phi$ and $d_{\rm min}$ are given for a
100\,AU circumstellar H disc with a temperature of 10\,000 K. For gas
temperatures of 5\,000K and 20\,000\,K the value of $\phi$ is
respectively higher and lower by $7.5\,\%$, while $d_{\rm min}$ is
$3.6\,\%$ larger and smaller, respectively.}   
\label{tab1}
\end{deluxetable}

\begin{figure}
\centering
\resizebox{\hsize}{!}{\includegraphics[width=16cm]{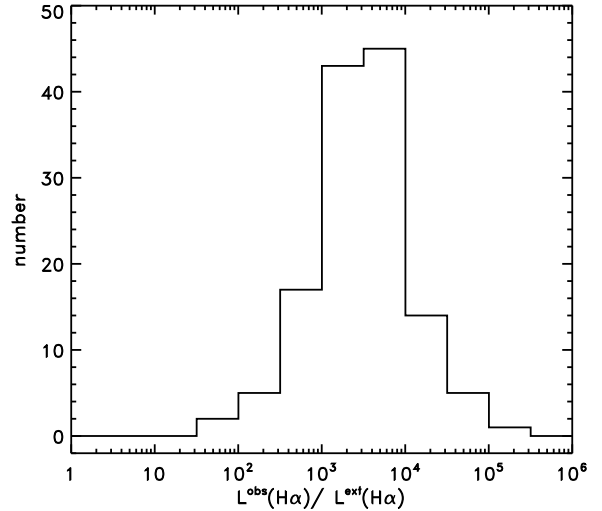}}
\caption{Histogram of the ratio of observed and external H$\alpha$
luminosity for our bona-fide PMS stars. The observed H$\alpha$
luminosity is typically three to four orders of magnitude larger than
the contribution due to external sources.}  
\label{fig12}
\end{figure}

It is convenient to calculate, for each PMS object, the
total H$\alpha$ luminosity $L^{\rm ext}(H\alpha)$ caused by all
neighbouring ionising stars, expressed as:

\begin{equation}
L^{\rm ext}(H\alpha) = \frac{r^2}{4} \, \frac{\alpha_{3,2}}{\beta_2} \, 
\sum_{i=1}^N \frac{(N_L)_i}{d_i^2}
\label{eq12}
\end{equation}

\noindent  
where $(N_L)_i$ is the ionising photon rate of star $i$ and $d_i$ is
its distance from the PMS object being considered. Besides the three
brightest stars mentioned above, we have included as potential sources
of the detected H$\alpha$ emission an additional 12 stars in the field
with $\log T_{\rm eff} \ge 4.4$ and $\log (L/L_\odot) \ge 4$,  as
classified by Romaniello et al. (2002). We note that these estimates of
$L^{\rm ext}(H\alpha)$ are firm upper limits because we are assuming
that {\em (i)} all disks are {\em face-on} relative to the hot stars,
and that {\em (ii)} the projected distances on the sky are indeed the
actual distances from the hot stars. The values of $L^{\rm ext}
(H\alpha)$ obtained in this way for the 133 bona-fide PMS stars span
the range  $10^{-6} - 10^{-4}$\,L$_\odot$, or typically three orders of
magnitude less than the median PMS H$\alpha$ luminosity, as shown by
the histogram in Figure\,\ref{fig12}. We can, therefore, conclude that
in this specific field the contribution of hot ionising stars to the
measured excess H$\alpha$ emission is negligible. We must stress,
however, that the effect might be much more pronounced in younger star
forming regions, such as the Orion Nebula or 30 Dor.

\subsection{Nebular continuum}

Finally, consideration must be given to the role of the nebular
continuum, i.e. radiation caused by bound--free and free--free
transitions in the gas and unrelated to the stellar photosphere. If
present, nebular continuum will add to the intrinsic continuum of the
star, thereby affecting both the observed total level and the slope.
This could ultimately alter the measured broad-band colours of the
source, thereby thwarting our attempts to infer the level of the
continuum in the H$\alpha$ band from the observed $V$ and $I$
magnitudes. It is important to estimate the magnitude of this effect,
as our method relies on the intrinsic $V-I$ colour of a star to derive
its $\Delta H\alpha$ (see Equation\,\ref{eq2}). 

Fortunately, the contribution of the nebular continuum appears to be
insignificant. To prove this, we have assumed a fully ionised gas of
pure H, considering only bound--free and free--free transitions and
ignoring the contribution to the continuum from two-photon emission
(Spitzer \& Greenstein 1951). Using Osterbrock's (1989, Chapter 4)
tabulations, we find H$\alpha$ line intensity and H$\alpha$ continuum
fluxes of the nebular gas as shown in Table\,\ref{tab2}, for gas
electron temperatures in the range 5\,000\,K -- 20\,000\,K. The purely
nebular $W_{\rm eq}(H\alpha)$ ranges from  $5\,000$\,\AA\,\, to
$9\,000$\,\AA, or two orders of magnitude higher than what we measure
for PMS objects. We can, therefore, safely conclude that the nebular
contribution to the continuum is negligible (less than 1\,\%).

Furthermore, we find that for gas temperatures in the range  5\,000\,K
-- 10\,000\,K the $V-I$ colour of the nebular continuum varies from
$1.4$ to $0.5$, spanning a range typical of G--K type stars. Thus the
effects of the nebular continuum on the $V-I$ colour of PMS stars
remains insignificant even for the objects with the  highest $W_{\rm
eq}(H\alpha)$ in our sample.

\begin{deluxetable}{rrrrr} 
\tablecolumns{5}
\tablewidth{0pc}
\tablecaption{Nebular emission and continuum of fully ionised H gas}
\tablehead{
\colhead{$T_{\rm el}$} & \colhead{$F_\nu$} & \colhead{$F_\lambda$} &
\colhead{$E(H\alpha)$} & \colhead{$W_{\rm eq}(H\alpha)$} }
\startdata
5\,000\,K & $10.28 \, 10^{-40} $ & $7.16 \, 10^{-29} $ & 
       $6.71 \, 10^{-25} $ & 9378\,\AA \\
10\,000\,K & $7.56 \, 10^{-40} $ & $5.26 \, 10^{-29} $ & 
       $3.56 \, 10^{-25} $ & 6763\,\AA \\
20\,000\,K & $5.51 \, 10^{-40} $ & $3.83 \, 10^{-29} $ & 
       $1.83 \, 10^{-25} $ & 4764\,\AA \\
\enddata
\tablecomments{$F_\nu$, $F_\lambda$ and $E(H\alpha)$ are given here in
erg cm$^3$  s$^{-1}$ Hz$^{-1}$, i.e. per unit volume ($V$), unit
electron density ($n_e$) and unit proton density ($n_p$). The
luminosity can be derived by multiplying the values listed here by
$V\,n_e\,n_p$.}   
\label{tab2}
\end{deluxetable}

\section{Discussion and conclusions}

We have developed and successfully tested a new self-consistent method
to reliably identify all PMS objects actively undergoing mass accretion
in a resolved stellar population, regardless of their age, without
requiring spectroscopy. The method combines broad-band $V$ and $I$
photometry with narrow-band $H\alpha$ imaging to: (1) identify all
stars with excess H$\alpha$ emission using as a reference template of
the photospheric luminosity in the H$\alpha$ band the average
$V-H\alpha$ colour of stars with small photometric errors; (2)
convert the excess H$\alpha$ magnitude into H$\alpha$ luminosity
$L(H\alpha)$ via the absolute photometric calibration of the H$\alpha$
band; (3) convert, if desired, the excess H$\alpha$ magnitude into the
H$\alpha$ emission equivalent width, using model spectra to fit the
level of the continuum in the H$\alpha$ band; (4) derive the accretion
luminosity $L_{\rm acc}$ from $L(H\alpha)$ using a relationship based
on recent literature values; (5) obtain the mass accretion rate $\dot
M_{\rm acc}$ from the free-fall equation that links it to $L_{\rm acc}$
via the stellar parameters (mass and radius), that we derive from the
H--R diagram using pre-main sequence tracks for the appropriate
metallicity.  

Since no spectroscopy of individual objects is needed, observations
requiring just a few hours of exposure time can suffice to reveal
hundreds of PMS stars simultaneously, determine their H$\alpha$
luminosity with an uncertainty of less than 15\,\% and derive their
mass accretion rates with the same accuracy as allowed by conventional
spectral line analysis.

As regards the H$\alpha$ luminosity, the virtue of this new method is
that it derives the reference luminosity of the stellar photosphere
inside the specific H$\alpha$ band simply from the average $V-H\alpha$
colour of stars with the same $V-I$ index (see Figure\,\ref{fig4}).
This is possible because, at any given time, the majority of the stars
in a stellar field have no excess H$\alpha$ emission. Obviously, the
average luminosity of the stellar photosphere in the H$\alpha$ band
reflects more or less pronounced H$\alpha$ absorption features,
depending on the spectral type of the objects, and precisely for this
reason it defines the reference luminosity with respect to which any
excess emission should be sought and measured. The uncertainty
associated with the derived $L(H\alpha)$ is dominated by statistical
errors in the photometry and does not exceed 15\,\% for our $4\,\sigma$
selection (see Section\,3.1). Note that this approach is also applicable  to
populations comprising mostly very young objects, such as the Orion
Nebula, since excess H$\alpha$ emission is found in only about one third
of the PMS stars at any given time and, therefore, the majority of the
very young objects will not have a conspicuous H$\alpha$  emission, thus
correctly defining the reference template (see Section\,3.1).

In spite of the good accuracy on $L(H\alpha)$, deriving the mass
accretion rate requires the use of a rather uncertain relationship
linking $L(H\alpha)$ to the accretion luminosity (see Section\,4 and
also Dahm 2008). The uncertainty on the relationship can have a 
systematic effect of up to a factor of 3 on the value of $\dot M_{\rm
acc}$. This is the unfortunate consequence of our incomplete
understanding of the physics of the accretion process, plagued by the
lack of direct measurements of the bolometric accretion luminosity
$L_{\rm acc}$, as well as by the paucity of simultaneous measurements
of the hot continuum excess emission and of the emission line fluxes in
H$\alpha$ or other lines (e.g. Pa$\beta$ and Br$\gamma$). Although
annoying, the systematic uncertainty of up to a factor of 3 on $\dot
M_{\rm acc}$ is similar to that plaguing age determinations based on
the comparison with theoretical isochrones, which can reach a factor
of 2 (see Section\,6.1).

On the other hand, these systematics affect in the same way all
determinations of $\dot M_{\rm acc}$ that make use of emission lines
diagnostics, including those based on a detailed study of the profile
and intensity of the H$\alpha$ line (e.g. Muzerolle et al. 1998b),
since their calibration relies on the same  measurements. For this
reason, the relative values of $\dot M_{\rm acc}$ that we  obtain are
trustworthy and the comparison of our results with those obtained with
other methods remains meaningful. 

As discussed in Sections\,3 and 4, as an application of our method, we
have studied the field of SN\,1987A, where we identify 133 bona-fide
PMS stars, based on their large H$\alpha$ excess. The median value of
the mass accretion rate that we find for these stars, namely $2.9
\times 10^{-8}$\,\Msolar\,yr$^{-1}$, is in very good agreement with the
median value of $2.4 \times 10^{-8}$\,\Msolar\,yr$^{-1}$ found by
Romaniello et al. (2004) from the analysis of the distribution of the
$(U-B)$ colour excess in the same field. While we have identified 133
bona-fide PMS stars with an H$\alpha$ excess above the $4\,\sigma$ level,
Romaniello et al. conclude that 765 stars with an excess $U$-band
luminosity larger than $\sim 0.035$\,L$_\odot$ could be bona-fide PMS
candidates, albeit at the $1.5\,\sigma$ level. 

\begin{figure}[t]
\centering
\resizebox{\hsize}{!}{\includegraphics[width=16cm]{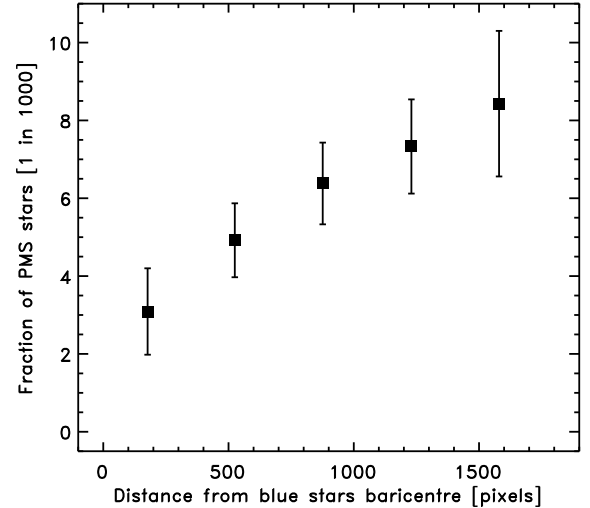}}
\caption{Fraction of PMS objects with respect to all stars, as a
function of the distance (in units of WFPC2 pixel or $0\farcs1$)
from the baricentre of the ionising stars.} 
\label{fig13}
\end{figure}

The approach followed by Romaniello et al. (2004) is statistical in
nature and, while providing a solid estimate of the average mass
accretion rate, it does not allow one to reliably identify individual
PMS stars. Therefore, Romaniello et al. turned to the H$\alpha$ excess
emission (namely $R-H\alpha$) as a more direct indicator of ongoing
accretion and found a strong correlation between the $U$-band continuum
excess and $W_{\rm eq}(H\alpha)$ for stars with $U-B > 0.5$ and $W_{\rm
eq}(H\alpha) > 3$\,\AA. Not surprisingly, the more stringent threshold
on the equivalent width adopted in the present study, i.e. $W_{\rm
eq}(H\alpha) > 10$\,\AA, results in a smaller number of identified of
bona-fide PMS stars, but also in a much smaller uncertainty. Indeed,
the method that we present here is more direct than the one based on
the $U$-band excess, since it does not require the use of model
atmospheres to predict the level of the intrinsic photospheric
continuum in the $U$ band from the observed optical colours, and as
such it results in a more reliable determination of the accretion
luminosity. 

Our method, therefore, makes it possible for the first time to carry
out a quantitative and systematic study of the properties of a large
number of PMS stars, including those in dense and/or distant star
forming regions. This allows us to tackle some still open questions as
to the evolution of both the stars and their discs. For instance, in
the field of SN\,1987A, we have studied how the mass accretion rate
changes with  time as our bona-fide PMS stars approach their main
sequence. We find that $\dot M_{\rm acc}$ decreases more slowly with
time than what is observed for low-mass T-Tauri stars in the Galaxy
(see  Section\,5) and predicted by models of viscous disc evolution
(Hartmann et al. 1998). However, our results are in excellent agreement
with the (so far limited) measurements of the mass accretion rate of G
type stars in the Milky Way (e.g. Sicilia--Aguilar et al. 2006), thus
lending credit to the hypothesis that $\dot M_{\rm acc}$ might depend
on the stellar mass. 

\begin{figure}[t]
\centering
\resizebox{\hsize}{!}{\includegraphics[width=16cm]{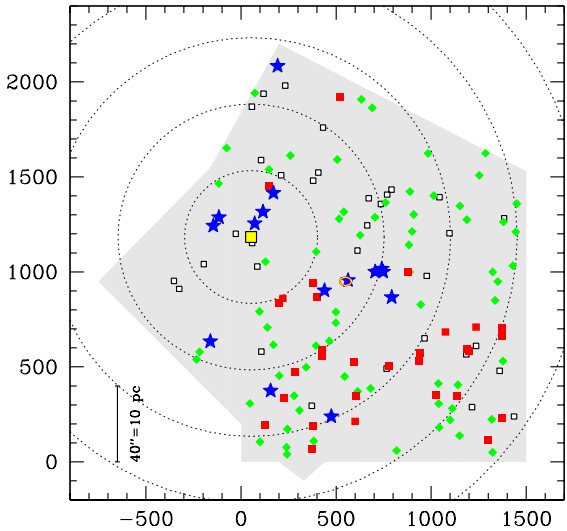}}
\caption{Map of the distribution of all bona-fide PMS objects with
respect to the ionising stars (blue star symbols), whose baricentre is
indicated by the large yellow square. Red squares correspond to stars
with $L(H\alpha) > 2 \times 10^{-2}$\,L$_\odot$, white squares indicate
objects with $L(H\alpha)< 8 \times 10^{-3}$\,L$_\odot$, and green
diamonds are used for intermediate values. Note the paucity of PMS
objects with high H$\alpha$ luminosity near the baricentre of the
ionising stars.} 
\label{fig14}
\end{figure}

Moreover, our method allows us to investigate whether and how the 
accretion process is affected by the chemical composition and density
of the molecular clouds or by the proximity of hot, massive ionising
stars. We will address the effects of metallicity in a forthcoming
paper (De Marchi et al., in preparation) devoted to a comparison
of star formation in the Small and Large Magellanic Clouds. As for the
effect of hot massive stars, we have studied the distribution of the
133 bona-fide PMS objects with respect to the baricentre of the 15
hottest stars in the field, described in Section\,6.3. We find a clear
anti-correlation, shown in Figure\,\ref{fig13}, between the frequency
of the identified PMS stars and their distance from the ionising
objects (marked with blue star symbols in Figure\,\ref{fig14}). This
effect is not seen for non-PMS objects of similar brightness, whose
distribution is uniform over the whole field, confirming that the trend
is not due to problems in detecting faint objects near the brightest
stars. The trend remains the same whether the baricentre of the hot
stars is computed using as weight their bolometric luminosity or their
ionising flux. This anti-correlation is quite remarkable in several
ways, as we discuss below. 

In Figure\,\ref{fig14} we show a schematic map of the distribution of
all 133 PMS objects with respect to the 15 ionising stars. The
baricentre of the ionising flux is shown by the large square and
concentric annuli are drawn from it, in increments of 35\arcsec. For
reference, a red ellipse marks the position of SN\,1987A. Symbols of
different colours indicate PMS stars with  different H$\alpha$
luminosity, from red for $L(H\alpha)> 2 \times 10^{-2}$\,L$_\odot$ to
white for $< 8 \times 10^{-3}$\,L$_\odot$, with green diamonds
corresponding to intermediate values. An inspection to
Figure\,\ref{fig14} immediately reveals two important facts: ({\em i})
there are very few PMS stars near the baricentre of the massive objects
and ({\em ii}) their H$\alpha$ luminosity is systematically lower than
that of PMS stars farther away. 

Panagia et al. (2000) had already pointed out that there is no 
overdensity of young PMS objects around hot massive stars in this
field. This shows that low-mass and massive stars do not follow each
other, suggesting that their formation mechanisms are separate and
distinct. But what Figure\,\ref{fig14} additionally shows is that 
there are very few PMS stars with strong H$\alpha$ luminosity (red
squares) around the young massive stars, with only 20\,\% of them
within a 70\,arcsec radius of the baricentre. Conversely, stars with
low H$\alpha$ luminosity (white squares) are distributed rather
uniformly across the field.

With a median age of $13.5$\,Myr, most PMS objects are much older than
the $1-2$\,Myr old hot stars and must belong to a previous generation.
Thus, no spatial relationship (whether a correlation or
anti-correlation) should be expected between the distribution of the
two types of objects. On the other hand, while we cannot exclude that
originally fewer PMS stars formed in the region now populated by young
massive stars, there are indications that the local paucity of PMS
objects is only apparent and caused by the presence of the hot young
stars. 

The fact that PMS objects near the youngest hot stars are both less
numerous and fainter in H$\alpha$ emission suggests that their
circumstellar discs have been considerably eroded and made less
efficient by enhanced photo-evaporation. We have shown earlier
(see Section\,6.3 and Figure\,\ref{fig12}) that the instantaneous
contribution to the H$\alpha$ luminosity from the ionising radiation  of
nearby massive stars is negligible. The results shown in
Figure\,\ref{fig14} confirms that this is the case. If it were not and
the observed H$\alpha$ luminosity were mostly due to fluorescence of
the circumstellar discs under the effect of the ionising radiation of
massive stars, the radial distribution should be opposite to the one
observed, with brighter H$\alpha$ luminosities for PMS stars closer to
the baricentre of the youngest hot stars. Thus, if the apparent paucity
of PMS objects in their vicinity is the result of photo-evaporation,
the latter must be mostly due to the effects of non-ionising far
ultraviolet (FUV) radiation. When integrated over the lifetime of
massive stars ($\sim 2$\,Myr), their FUV radiation can be quite
significant for the disruption of nearby circumstellar discs.
In particular, this also means
that there might be many more objects still in their PMS phase, close
to the hot stars, but we have no way to identify them as such, since
the photo-evaporation of their discs has stopped the accretion process.

Recent theoretical studies of the erosion of circumstellar discs via
photo-evaporation due to external FUV radiation  (e.g. Adams et al. 
2004) or to the PMS star itself (e.g. Alexander, Clarke \&
Pringle 2006a; 2006b) suggest timescales of order 8--10\,Myr for the
complete dissolution of the discs. Although this is the right order of
magnitude for the timescale of the process that we see, there are still
some discrepancies. For instance, it is clear that if discs were to
completely dissolve under the effect of their own central star within
5--10\,Myr, as suggested by Alexander et al. (2006b), the majority of our
PMS objects, with a median age of $13.5$\,Myr, should not show any sign
of accretion. Similarly, with a typical age of 2\,Myr, the hottest
stars in our field would have not had time to deposit enough UV photons
on to the PMS discs to seriously erode them, if the process is to take
of order 10\,Myr (Adams et al. 2004). While recently Clarke (2007)
suggested that the effect of nearby massive young stars may be
stronger, leading to disc exhaustion timescales of order 2\,Myr, her
calculations apply to the immediate vicinity ($< 1$\,pc) of very 
massive stars, such as the O6 star $\theta_1 C$ in Orion. The density
of massive stars in the field of SN\,1987A is much lower than that
present in the Orion Nebula and the distances to our PMS objects
typically an order of magnitude larger than those considered by Clarke
(2007). 

Nevertheless, while it is clear that the theory of disc
photo-evaporation does not yet capture all the details, we believe that
it can greatly benefit from the observations presented here. In
particular, it will be necessary to explain which mechanisms allow
isolated circumstellar discs  to feed their central PMS stars for well
over 20\,Myr, at least in a low-metallicity, low-density environment
such as the field of the LMC. At the same time, the models should
justify the quick ($\lesssim 3$\,Myr) demise of the PMS discs within a
(projected) radius of $\sim 10$\,pc of a handful of late O stars.
Extending our investigation, as we plan to do, to more and denser
regions, such as NGC\,1850 and 30\,Dor in the Large Magellanic Cloud
and NGC\,346 and NGC\,602 in the Small Magellanic Cloud, will offer
increased statistics and a much wider variety of star formation
environments.

\begin{acknowledgements}

We are very thankful to Pier Giorgio Prada Moroni and Scilla
Degl'Innocenti for sharing with us their pre-main sequence evolutionary
tracks ahead of publication and to Massimo Robberto for useful
discussions. GDM is grateful to STScI and ESO for the hospitality, via
their Science Visitor programmes, during the preparation of this paper.
The research of NP was supported in part by NASA grant HST-GO-11547.06-A. 
We are also indebted to an anonymous referee for precious comments
and suggestions that have helped us to improve the presentation of our
work.

\end{acknowledgements}


\section*{Appendix}

\begin{deluxetable*}{lclll} 
\tablecolumns{5}
\tablewidth{0pc}
\tablecaption{Predicted colour relations between $R$ or $H\alpha^c$ and
$V-I$ for various photometric systems.}
\tablehead{
\colhead{System} & \colhead{Band} & \colhead{BPGS, $[M/H]=0$} & 
\colhead{Bessell et al., $[M/H]=-0.5$}  &
\colhead{Bessell et al., $[M/H]=-1$} }
\startdata
 WFPC2 &  $m_{675}$ & 
 $0.342 \times m_{555} + 0.658 \times m_{814}$ & 
 $0.354 \times m_{555} + 0.646 \times m_{814}$ &
 $0.353 \times m_{555} + 0.647 \times m_{814}$ \\
       &  $m_{675}$ & 
 $0.510 \times m_{606} + 0.490 \times m_{814}$ & 
 $0.510 \times m_{606} + 0.490 \times m_{814}$ & 
 $0.508 \times m_{606} + 0.492 \times m_{814}$ \\
       &  $m_{656}^c$ &  & 
 $0.409 \times m_{555} + 0.591 \times m_{814} - 0.351$ & 
 $0.411 \times m_{555} + 0.589 \times m_{814} - 0.350$ \\
       &  $m_{656}^c$ &  & 
 $0.593 \times m_{606} + 0.407 \times m_{814} - 0.353$ & 
 $0.595 \times m_{606} + 0.405 \times m_{814} - 0.351$ \\
\cline{1-5} 
 ACS/ &  $m_{625}$ &
 $0.496 \times m_{555} + 0.504 \times m_{814}$ & 
 $0.499 \times m_{555} + 0.501 \times m_{814}$ &
 $0.503 \times m_{555} + 0.497 \times m_{814}$ \\
 WFC     &  $m_{625}$ &
 $0.708 \times m_{606} + 0.292 \times m_{814}$ & 
 $0.703 \times m_{606} + 0.297 \times m_{814}$ & 
 $0.708 \times m_{606} + 0.292 \times m_{814}$ \\
         &  $m_{658}^c$ &  & 
 $0.381 \times m_{555} + 0.619 \times m_{814} - 0.156$ & 
 $0.386 \times m_{555} + 0.614 \times m_{814} - 0.158$ \\
         &  $m_{658}^c$ &  & 
 $0.546 \times m_{606} + 0.454 \times m_{814} - 0.160$ & 
 $0.552 \times m_{606} + 0.448 \times m_{814} - 0.164$ \\
\hline
 ACS/ &  $m_{625}$ &
 $0.506 \times m_{555} + 0.494 \times m_{814}$ & 
 $0.502 \times m_{555} + 0.498 \times m_{814}$ &
 $0.511 \times m_{555} + 0.489 \times m_{814}$ \\
 HRC     &  $m_{625}$ &
 $0.707 \times m_{606} + 0.293 \times m_{814}$ & 
 $0.697 \times m_{606} + 0.303 \times m_{814}$ & 
 $0.704 \times m_{606} + 0.296 \times m_{814}$ \\
         &  $m_{658}^c$ &  & 
 $0.385 \times m_{555} + 0.615 \times m_{814} - 0.160$ & 
 $0.390 \times m_{555} + 0.610 \times m_{814} - 0.160$ \\
         &  $m_{658}^c$ &  & 
 $0.542 \times m_{606} + 0.458 \times m_{814} - 0.166$ & 
 $0.544 \times m_{606} + 0.456 \times m_{814} - 0.165$ \\
\hline
 WFC3/    &  $m_{625}$ &
 $0.504 \times m_{555} + 0.496 \times m_{814}$ & 
 $0.510 \times m_{555} + 0.490 \times m_{814}$ &
 $0.517 \times m_{555} + 0.483 \times m_{814}$ \\
 UVIS     &  $m_{625}$ &
 $0.742 \times m_{606} + 0.258 \times m_{814}$ & 
 $0.734 \times m_{606} + 0.266 \times m_{814}$ & 
 $0.741 \times m_{606} + 0.259 \times m_{814}$ \\
           &  $m_{656}^c$ &  & 
 $0.387 \times m_{555} + 0.613 \times m_{814} - 0.405$ & 
 $0.389 \times m_{555} + 0.611 \times m_{814} - 0.408$ \\
           &  $m_{656}^c$ &  & 
 $0.558 \times m_{606} + 0.442 \times m_{814} - 0.406$ & 
 $0.560 \times m_{606} + 0.440 \times m_{814} - 0.409$ \\
\hline
 Johnson- &  $R$ &  
 $0.451 \times V + 0.549 \times I$ &
 $0.465 \times V + 0.535 \times I$ &
 $0.473 \times V + 0.527 \times I$ \\
 Cousin   & & & & 
\enddata
\tablecomments{The typical $1\,\sigma$ uncertainty on the slope of the
regression fit is $0.003$.}   
\label{tab3}
\end{deluxetable*}

Over the years, a large number of observations of stellar populations 
have been collected with the HST/WFPC2 and HST/ACS through the $V$-like
filters F555W and F606W and the $I$-like filter F814W. We have
therefore computed colour relationships between the magnitude in the
$R$-like band F675W ($m_{675}$) and the $V-I$-like colours
$m_{555}-m_{814}$ and $m_{606}-m_{814}$. To this aim, we have used both
observed and theoretical stellar spectra, taken respectively from the
Bruzual-Persson-Gunn-Stryker (BPGS) Spectrophotometry Atlas (see Gunn
\& Stryker 1983) and from the stellar atmosphere models of Bessell et
al. (1998). 

The BPGS atlas contains nearby bright dwarfs and giants of near-solar
metallicity, whereas from the library of Bessell et al. (1998) we have
selected specifically those models with metallicity index $[M/H]=-0.5$,
as appropriate for the LMC (Dufour 1984), surface gravity $g=4.5$
and effective temperatures in the range 3\,500 \,K $\le T_{\rm
eff} \le $ 40\,000\,K. Using the standard HST synthetic
photometry package Synphot (Laidler et al. 2008), we have folded these
spectra through the instrumental response of the WFPC2, the ACS/WFC,
the ACS/HRC and the WFC3/UVIS cameras, in order to obtain the expected
magnitudes in the bands listed above. For comparison with ground-based
photometry, we have also calculated the expected magnitudes through the
Johnsn--Cousin $V, R$ and $I$ bands. 

\begin{figure}
\centering
\resizebox{\hsize}{!}{\includegraphics[width=16cm]{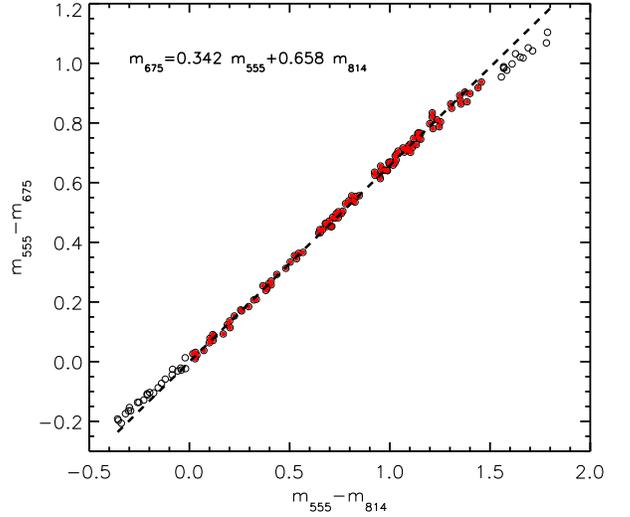}}
\caption{Relationship between the $m_{555}-m_{675}$ and
$m_{555}-m_{814}$ colours of the WFPC2 filter system based on all  the
observed spectra in the BPGS atlas, with solar metallicity. The range
over which the least square fit is carried out is indicated by filled
circles.} 
\label{fig15}
\end{figure}

\begin{figure}
\centering
\resizebox{\hsize}{!}{\includegraphics[width=16cm]{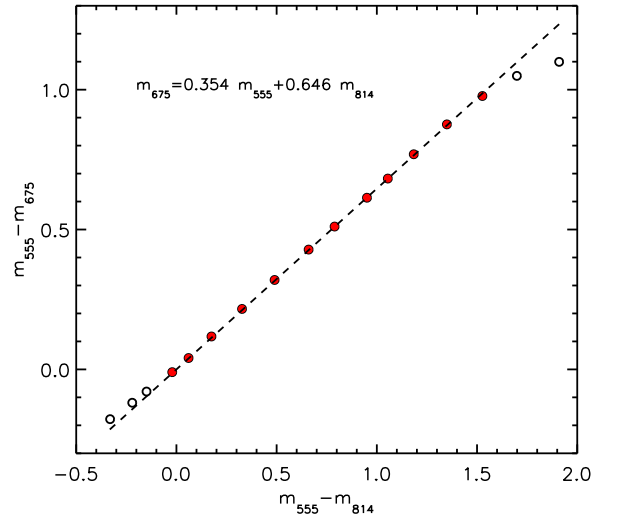}}
\caption{Relationship between the $m_{555}-m_{675}$ and
$m_{555}-m_{814}$ colours of the WFPC2 filter system based on model
spectra of Bessell et al. (1998) for metallicity $[M/H]=-0.5$. The
range over which the least square fit is carried out (4\,000\,K $<
T_{\rm eff} <$ 10\,000\,K) is indicated by filled circles.} 
\label{fig16}
\end{figure}

We show in Figure\,\ref{fig15}, for the WFPC2 filter system, the
predicted  $m_{555}-m_{675}$ colour as a function of $m_{555}-m_{814}$
for all the stars in the BPGS atlas. The distribution is remarkably
linear and the dashed line represents the best fit in the colour range
$0 < m_{555}-m_{814} < 1.5$ (filled symbols), corresponding to the
relationship $m_{675} = 0.34 \times m_{555} + 0.66 \times m_{814}$. The
colour--colour diagram in the same bands based on the models of Bessell
et al. (1998) is shown in Figure\,\ref{fig16} and the relationship is
practically indistinguishable, namely $m_{675} = 0.35 \times m_{555} +
0.65 \times m_{814}$. In this case, the dashed line shows the
least-square fit to stars in the range 4\,000\,K $\le T_{\rm
eff} \le$ 10\,000\,K, marked by the filled symbols. As already
noted in Section\,2, it appears that neither the luminosity class nor
the metallicity has a noticeable impact on the colour relationship in
these bands.

\begin{figure}
\centering
\resizebox{\hsize}{!}{\includegraphics[width=16cm]{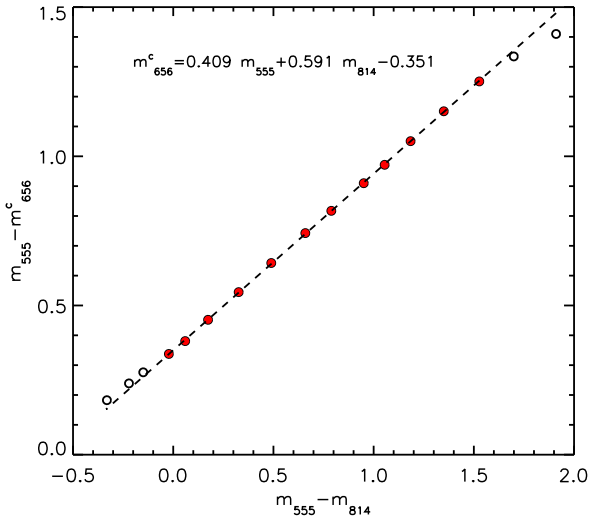}}
\caption{Relationship between the $m_{555}-m_{656}^c$ and
$m_{555}-m_{814}$ colours of the WFPC2 filter system based on model
spectra of Bessell et al. (1998), for metallicity $[M/H]=-0.5$, in
which all spectral features in the range 6\,500 -- 6\,620\,\AA\,\, have
been replaced by the continuum. The range over which the least square
fit is carried out (4\,000\,K $< T_{\rm eff} <$ 10\,000\,K) is
indicated by filled circles.} 
\label{fig17}
\end{figure}

\begin{figure}
\centering
\resizebox{\hsize}{!}{\includegraphics[width=16cm]{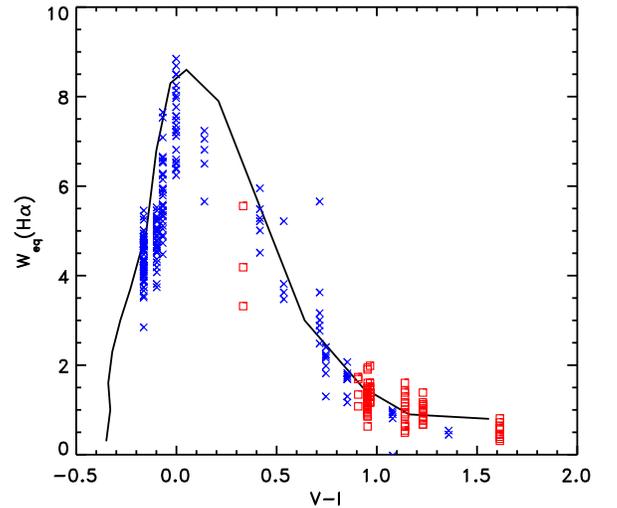}}
\caption{Comparison between the H$\alpha$ absorption equivalent widths
in the catalogue of Ducati (1981; crosses indicate dwarfs, squares are
for giants) and those derived by applying Equation\,\ref{eq4} to the
spectral models of Munari et al. (2005) for solar metallicity (solid
line).} 
\label{fig18}
\end{figure}

This is also the case when we consider the $m_{606}-m_{814}$ colour,
for which the relationship with $m_{675}$ is given in
Table\,\ref{tab3}. In the same table we also list the results for the
ACS/WFC, ACS/HRC and WFC3/UVIS cameras, as well as those for the
standard Johnson--Cousin system normally used from the ground (note
that the $R$ and $I$ bands in this system do not coincide with the $R$
and $I$ bands of Johnson's original system, as mentioned in
Section\,2). As expected, the coefficients in the colour relationship
vary considerably with the band, and differences exist also between the
same band on various HST cameras (or, at least, bands with the same
name).  

As briefly discussed in Section\,3, the magnitude corresponding to the
level of the continuum in the H$\alpha$ band can be derived by means of
properly validated model atmospheres, such as those of Bessell et al.
(1998) that we have shown to reproduce quite accurately the observed
broad-band colours (see Figure\,\ref{fig2}). This means that, even
though the models might lack the spectral resolution necessary to
realistically reproduce spectral lines (hence the small discrepancy
between the dashed lines and the squares in Figure\,\ref{fig2}), the
level of the continuum can be trusted. We have, therefore, fitted the
continuum in the Bessell et al. (1998) spectra over the range $6\,500 -
6\,620$\,\AA\, and have replaced the actual spectrum in that range with
the interpolated continuum. Finally, we have used Synphot to fold the
model spectra, modified in this way, through the instrumental set-up
and derive in this way the continuum H$\alpha$ magnitudes ($H\alpha^c$)
for the various H$\alpha$ filters on-board the HST, as a function of
the  spectral type or $T_{\rm eff}$ of the star.

With this information it is now possible to build colour--colour
diagrams like those of Figure\,\ref{fig15} and \ref{fig16}, and look
for a colour relationship between $H\alpha^c$ and $V-I$. In
Figure\,\ref{fig17} we show the specific case of the WFPC2. The
$m_{555}-m^c_{656}$ colour varies very smoothly and very slowly with
the  effective temperature and is a remarkably linear function of the
$m_{555}-m_{814}$ colour over the whole range, and particularly for
stars with 4\,000\,K $< T_{\rm eff} < 10\,000$\,K (filled circles) with
which we are mostly concerned in this work. Similar relationships exist
for other bands and photometric systems and they are given in
Table\,\ref{tab3}. We note that the constant $0.351$ in the colour
relationship accounts for the conspicuous H$\alpha$ absorption line in
the spectrum of Vega, which defines the standard for the VEGAMAG
magnitude system.


As indicated in Section\,3, the difference between the observed
$H\alpha$ magnitude and $H\alpha^c$ can be converted to the equivalent
width $E_{\rm eq}(H\alpha)$ by means of Equation\,\ref{eq4}. To prove
this, we have applied Equation\,\ref{eq4} to the model  atmospheres of
Munari et al. (2005), which provide spectra with a resolving power of
$11\,500$ and a 1\,\AA\, sampling, and compared the results to
literature measurements of H$\alpha$ equivalent widths (in absorption).

Ducati (1981) compiled an extensive homogeneous catalogue of H$\alpha$
line measurements, from photographic and photoelectric techniques, for
about 2\,300 stars. We have converted Ducati's index $\alpha$ to
$W_{\rm eq}($H$\alpha)$ following the calibration provided by Strauss
\& Ducati (1981) and Cester et al. (1977). The values of $W_{\rm
eq}($H$\alpha)$ obtained in this way are shown in Figure\,\ref{fig18},
where crosses correspond to dwarfs and boxes to giants. Ducati's (1981)
stars are classified according to their spectral type, which we have
converted to $V-I$ following Bessell (1990). 

As for the application of Equation\,\ref{eq4} to the models of Munari
et al. (2005), this was done for model spectra with $4\,000 < T_{\rm
eff}/{\rm K} <40\,000$,  $\log g =4.5$ and solar metallicity and for
the WFPC2 photometric system, namely F555W for V, F814W for I and F656N
for H$\alpha$. It should be noted that, for stars with broad H$\alpha$
lines (around spectral type A), the equivalent width determined in this
way is systematically smaller than the true one, since part of the line
wings fall outside of the H$\alpha$ filter. However, the measurements
of Ducati (1981) are subject to exactly the same effect, since they
were derived in a similar fashion and with a H$\alpha$ filter of
comparable effective width. The value of $W_{\rm eq}($H$\alpha)$
obtained in this way is shown in Figure\,\ref{fig18} (solid line), as a
function of $V-I$. The agreement with the observations is very
satisfactory.

\begin{deluxetable}{lccc} 
\tablecolumns{4}
\tablewidth{235pt}
\tablecaption{Rectangular width and reference wavelength of past and
present H$\alpha$ filters installed on board the HST.}
\tablehead{
\colhead{Instrument} & \colhead{Filter} & \colhead{$RW$ [\AA]} & 
\colhead{$\lambda_{\rm ref}$} }
\startdata
WFPC2   & F656N  &  $28.35$ & $6563.8$\\
ACS/WFC & F658N  &  $74.96$ & $6584.3$\\
ACS/HRC & F658N  &  $74.96$ & $6583.9$\\
WFC3    & F656N  &  $17.68$ & $6561.4$\\
        & F658N  &  $27.61$ & $6585.0$
\enddata
\label{tab4}
\end{deluxetable}



\begin{thebibliography}{References}
 
\bibitem[]{} Adams, F., Hollenbach, D., Laughlin, G., Gorti, U. 2004,
ApJ, 611, 360 

\bibitem[]{} Alencar, S., Johns--Krull, C., Basri, G. 2001, AJ,
122, 3335

\bibitem[]{} Alexander, R. D., Clarke, C. J., Pringle, J. E. 2006a,
MNRAS, 369, 216 

\bibitem[]{} Alexander, R. D., Clarke, C. J., Pringle, J. E. 2006b,
MNRAS, 369, 229 

\bibitem[]{} Appenzeller, I., Mundt, R. 1989, A\&ARv, 1, 291 

\bibitem[]{} Baker, J., Menzel, D. 1938, ApJ, 88, 52

\bibitem[]{} Bertout, C. 1989, ARA\&A, 27, 351

\bibitem[]{} Bessell, M. 1990, A\&AS, 83, 357 

\bibitem[]{} Bessell, M., Castelli, F., Plez, B. 1998, A\&A, 333, 231

\bibitem[]{} Calvet, N., Hartmann, L., Strom, E. 2000, in ``Protostars and
  Planets'', eds V. Mannings, A. Boss, S. Russell (Tucson: University of
  Arizona Press), 377

\bibitem[]{} Calvet, N., Muzerolle, J., Brice\~no, C., Hernandez, J.,
Hartmann, L., Saucedo, J. L., Gordon, K. D. 2004, AJ, 128, 1294

\bibitem[]{} Cester, B., Giuricin, G., Mardirossian, F., Pucillo, M.,
Castelli, F.,  Flora, U. 1977, A\&AS, 30, 1 

\bibitem[]{} Chieffi, A., Straniero, O. 1989, ApJS, 71, 47

\bibitem[]{} Cignoni, M., Sabbi, E., Nota, A., Tosi, M.,
Degl'Innocenti, S., Prada Moroni, P.G., Angeretti, L., Carlson, L.R.,
Gallagher, J., Meixner, M., Sirianni, M., Smith, L. 2009, AJ, 137, 3668

\bibitem[]{} Clarke, C. J. 2007, MNRAS, 376, 1350 

\bibitem[]{} Clarke, C., Pringle, J. 2006, MNRAS, 370, L10

\bibitem[]{} Dahm, S. 2008, AJ, 136, 521

\bibitem[]{} D'Antona, F., Mazzitelli, I. 1997, MmSAI, 68, 807 

\bibitem[]{} Degl'Innocenti, S., Prada Moroni, P.G., Marconi, M.,
Ruoppo, A. 2008, Ap\&SS, 316, 25

\bibitem[]{} Drew, J., et al. 2005, MNRAS, 362, 753

\bibitem[]{} Ducati, J. 1981, A\&AS, 45, 119 

\bibitem[]{} Edwards, S., Cabrit, S., Strom, S., Heyer, I., Strom, K.,
Anderson, E. 1987, ApJ, 321, 473

\bibitem[]{} Fernandez, M., Ortiz, E., Eiroa, C., Miranda, L.  1995,
A\&AS, 114, 439

\bibitem[]{} Geha, M. et al. 1998, AJ, 115, 1045 

\bibitem[]{} Gilmozzi, R., Kinney, E., Ewald, S., Panagia, N.,
Romaniello, M. 1994, ApJ, 435, L43 

\bibitem[]{} Gouliermis, D. A., Henning, T., Brandner, W., Dolphin, A.
E., Rosa, M.,  Brandl, B. 2007, ApJ, 665, L27 

\bibitem[]{} Gullbring, E., Hartmann, L., Brice\~no, C., Calvet, N.
1998, ApJ, 492, 323

\bibitem[]{} Gunn, J., Stryker, L. 1983, ApJS, 52, 121 

\bibitem[]{} Hartigan, P., Edwards, S., Ghandour, L. 1995, ApJ, 452, 736

\bibitem[]{} Hartmann, L., Calvet, P., Gullbring, E., D'Alessio, P.
1998, ApJ, 495, 385

\bibitem[]{} Herbig, G. 1957, ApJ, 125, 654

\bibitem[]{} Heyer, I., Biretta, J. 2004, ``WFPC2 Instrument Handbook'',
(Baltimore: STScI)

\bibitem[]{} Hill, V., Andrievsky, S., Spite, M. 1995, A\&A, 293, 347

\bibitem[]{} Horne, K. 1988, in ``New Directions in Spectrophotometry'',
eds, A. Davis Philip, D. Hayes, S. Adelman (Schenectady: Davis Press), 145

\bibitem[]{} Hunter, D., Shaya, E., Holtzman, J., Light, R., O'Neil, E., 
Lynds, R. 1995, ApJ, 448, 179 

\bibitem[]{} Johnson, H. L. 1966, ARA\&A, 4, 193  

\bibitem[]{} K\"onigl, A. 1991, ApJ, 370, L39

\bibitem[]{} Koornneef, J., Bohlin, R., Buser, R., Horne, K., Turnshek,
D. 1986, in ``Highlights of astronomy'', Volume 7, (Dordrecht:
Reidel),  833

\bibitem[]{} Kohoutek, L., Wehmeyer, R. 1999, A\&AS, 134, 255

\bibitem[]{} Laidler, V. et al. 2008, ``Synphot Data User's Guide''
(Baltimore: STScI)

\bibitem[]{} Lynden--Bell, D., Pringle, J. 1974, MNRAS, 168, 603

\bibitem[]{} Munari, U., Sordo, R., Castelli, F., Zwitter, T. 2005,
A\&A, 442, 1127 

\bibitem[]{} Muzerolle, J., Hartmann, L., Calvet, N., 1998a, AJ, 116,
455 

\bibitem[]{} Muzerolle, J., Calvet, N.,  Hartmann, L., 1998b, ApJ, 492,
743

\bibitem[]{} Muzerolle, J., Calvet, N., Brice\~no, C., Hartmann, L.,
Hillenbrand, L. 2000,  ApJ, 535, L47

\bibitem[]{} Muzerolle, J., Hillenbrand, L., Calvet, N., Brice\~no, C., 
Hartmann, L., 2003, ApJ, 592,266

\bibitem[]{} Muzerolle, J., Luhman, K., Brice\~no, C., Hartmann, L.,
  Calvet, N. 2005, ApJ, 625, 906

\bibitem[]{} Natta, A., Testi, L., Muzerolle, J., Randich, S., et al.
2004, A\&A, 424, 603 

\bibitem[]{} Natta, A., Testi, L., Randich, S. 2006, A\&A, 452, 245 

\bibitem[]{} Nota, A., Sirianni, M., Sabbi, E., Tosi, M., Clampin, M., 
Gallagher, J., Meixner, M., Oey, M. S., Pasquali, A., Smith, L. J., 
Walterbos, R., Mack, J. 2006, ApJ, 640, L29

\bibitem[]{} Osterbrock, D. 1989, ``Astrophysics of gaseous nebulae and 
active galactic nuclei'', (Mill Valley: University Science Books)

\bibitem[]{} Padgett, D. 1996, ApJ, 471, 847

\bibitem[]{} Palla, F., Stahler, S. 1993, ApJ, 418, 414 

\bibitem[]{} Panagia, N. 1973, AJ, 78, 929

\bibitem[]{} Panagia, N. 1999, in ``New Views of the Magellanic
Clouds'', IAU Symp. 190, eds. Y.-H. Chu, N. Suntzeff, J. Hesser, D.
Bohlender (San Francisco: ASP), 549

\bibitem[]{} Panagia, N., Gilmozzi, R., Macchetto, F., Adorf, H.-M.,
Kirshner, R. P.  1991, ApJ, 380, L23

\bibitem[]{} Panagia, N., Romaniello, M., Scuderi, S., Kirshner, R. 2000,
ApJ, 539, 197

\bibitem[]{} Parker, Q., et al. 2005, MNRAS, 362, 689

\bibitem[]{} Pasquini, L., de Medeiros, J., Girardi, L. 2000, A\&A,
361, 1011

\bibitem[]{} Romaniello, M. 1998, PhD thesis, Scuola Normale Superiore,
Pisa, Italy

\bibitem[]{} Romaniello, M., Panagia, N., Scuderi, S., Kirshner, R. 2002,
AJ, 123, 915

\bibitem[]{} Romaniello, M., Robberto, M., Panagia, N. 2004, ApJ, 608,
220

\bibitem[]{} Romaniello, M., Scuderi, S., Panagia, N., Salerno, R. M.,
Blanco, C. 2006, A\&A, 446, 955 

\bibitem[]{} Scuderi, S., Panagia, N., Gilmozzi, R., Challis, P.,
Kirshner, R.  1996, ApJ, 465, 956 

\bibitem[]{} Sicilia-Aguilar, A., Hartmann, L., Hernandez, J.,
Brice\~no, C., Calvet,  N. 2005, AJ, 130, 188 

\bibitem[]{} Sicilia--Aguilar, A., Hartmann, L., Furesz, G., Henning,
T., Dullemond, C., Brandner, W. 2006, AJ, 132, 2135 

\bibitem[]{} Siess, L., Dufour, E., Forestini, M. 2000, A\&A, 358, 593

\bibitem[]{} Sirianni, M., Jee, M., Benitez, N., Blakeslee, J., Martel, A., 
Meurer, G., Clampin, M., De Marchi, G., Ford, H. C., Gilliland, R., Hartig, 
G., Illingworth, G., Mack, J., McCann, W. 2005, PASP, 117, 1049

\bibitem[]{} Sirianni, M., Nota, A., Leitherer, C., De Marchi, G.,
Clampin, M. 2000,  ApJ, 533, 203 

\bibitem[]{} Shu, F., Najita, J., Ruden, S., Lizano, S. 1994, ApJ, 429, 727

\bibitem[]{} Smith, K., Lewis, G., Bonnell, I., Bunclark, P., Emerson,
J. 1999, MNRAS, 304, 367

\bibitem[]{} Spitzer, L., Greenstein, J. 1951, ApJ, 114, 407 

\bibitem[]{} Strauss, F., Ducati, J. 1981, A\&AS, 44, 337 

\bibitem[]{} Ulmschneider, P. 1979, Space Sci. Rev., 24, 71

\bibitem[]{} von Braun, K., Chiboucas, K., Minske, J. K., Salgado, J.
F., Worthey, G.  1998, PASP, 110, 810

\bibitem[]{} White, R., Basri, G. 2003, ApJ, 582, 1109

\bibitem[]{} White, R., Hillenbrand, L. 2004, ApJ, 616, 998


\end{thebibliography}
\end{document}